\documentclass[aps,prl,twocolumn,superscriptaddress,showpacs,english]{revtex4-1}

\usepackage[T1]{fontenc}
\usepackage{lipsum}

\usepackage[latin9]{inputenc}
\usepackage{babel}
\usepackage{amsmath}
\usepackage{amssymb}
\usepackage{wasysym}
\usepackage{graphicx}
\usepackage{multirow}
\usepackage{gensymb}
\usepackage[dvipsnames]{xcolor}
\usepackage{pifont}
\usepackage{microtype}
\usepackage{xcolor}
\usepackage{soul}
\usepackage[version=3]{mhchem}
\usepackage{chemformula}[2014/04/08] %nice x's
\usepackage{booktabs}    % cmdirule
\usepackage{dcolumn}     % Align table columns on decimal point
\usepackage{bm}          % bold math
\usepackage[linktocpage=true,
  colorlinks=true, 
  pdfborder={0 0 0},
  linkcolor=blue,
  citecolor=red,
  filecolor=yellow,
  urlcolor=blue,
  bookmarks,
  pdfauthor={},
]{hyperref}

\usepackage{etoolbox} % Provides ability to hook into thebibliography
\newcounter{firstbib} % New counter: keeps track of the next first number 

\AtBeginEnvironment{thebibliography}{
}

\apptocmd{\thebibliography}{
  \setcounter{NAT@ctr}{\value{firstbib}} % Use this for natbib and revtex
}{}{}  % These args are for errors handling: see etoolbox docs

\AtEndEnvironment{thebibliography}{
  \setcounter{firstbib}{\value{NAT@ctr}}
}

\newcommand{\laH}{LaH$_{10}$}
\newcommand{\laD}{LaD$_{10}$}

\newcommand{\sh}{H$_3$S}

\newcommand{\tc}{T$_{\text{c}}$}

\newcommand{\omlog}{$\omega_{\text log}$}
\newcommand{\tcscdft}{T$_{\text{c}}^{\rm SCDFT}$}

\newcommand{\fcc}{$Fm$-$3m$}

\newcommand{\Unidonostia}{Fisika Aplikatua 1 Saila, Gipuzkoako Ingeniaritza Eskola, University of the Basque Country (UPV/EHU), Europa Plaza 1, 20018 Donostia/San Sebasti\'an, Spain}
\newcommand{\CFM}{Centro de F\'isica de Materiales (CSIC-UPV/EHU), Manuel de Lardizabal Pasealekua 5, 20018 Donostia/San Sebasti\'an, Spain}
\newcommand{\DIPC}{Donostia International Physics Center (DIPC),  Manuel de Lardizabal Pasealekua 4, 20018 Donostia/San Sebasti\'an, Spain}
 
\newcommand{\MPIHalle}{Max-Planck Institute of Microstructure Physics, Weinberg 2, 06120 Halle, Germany}
\newcommand{\UTokyo}{Department of Applied Physics, University of Tokyo, 7-3-1 Hongo Bunkyo-ku, Tokyo 113-8656 Japan}
\newcommand{\USendai}{Department of Physics, Tohoku University, 6-3 Aza-Aoba, Sendai, 980-8578 Japan}
\newcommand{\RIKEN}{RIKEN Center for Emergent Matter Science, 2-1 Hirosawa, Wako, 351-0198, Japan}
\newcommand{\Tsukuba}{Research Center for Magnetic and Spintronic Materials, National Institute for Materials Science, Tsukuba 305-0047, Japan}
\newcommand{\UniParis}{Sorbonne Universit\'e, CNRS, Institut des Nanosciences de Paris, UMR7588, F-75252, Paris, France}
\newcommand{\UniRoma}{Dipartimento di Fisica, Universit\`a di Roma La Sapienza, Piazzale Aldo Moro 5, I-00185 Roma, Italy}
\newcommand{\graphene}{Graphene Labs, Fondazione Istituto Italiano di Tecnologia, Via Morego, I-16163 Genova, Italy}

\begin{document}
%===============
% \begin{bibunit}[plain]
\title{Quantum Crystal Structure in the 250 K Superconducting Lanthanum Hydride}  

% back up version on 26.07.2019; This is master version.    j 

\author{Ion Errea}              \affiliation{\Unidonostia} \affiliation{\CFM} \affiliation{\DIPC}
\author{Francesco Belli}        \affiliation{\Unidonostia} \affiliation{\CFM}
\author{Lorenzo Monacelli}      \affiliation{\UniRoma}
\author{Antonio Sanna}          \affiliation{\MPIHalle}
\author{Takashi Koretsune}      \affiliation{\USendai}
\author{Terumasa Tadano}        \affiliation{\Tsukuba}
\author{Raffaello Bianco}       \affiliation{\CFM}
\author{Matteo Calandra}        \affiliation{\UniParis}
\author{Ryotaro Arita}          \affiliation{\UTokyo}\affiliation{\RIKEN}
\author{Francesco Mauri}        \affiliation{\UniRoma} \affiliation{\graphene}
\author{Jos\'e~A. Flores-Livas} \affiliation{\UniRoma} 

\date{\today}

\begin{abstract}
The discovery of superconductivity at 200\,K in the hydrogen sulfide system at large pressures~\cite{DrozdovEremets_Nature2015} 
was a clear demonstration that hydrogen-rich materials can be high-temperature superconductors. 
The recent synthesis of \laH\ with a superconducting critical temperature (\tc) 
of 250\,K~\cite{Hemley-LaH10_PRL_2019,Nature_LaH_Eremets_2019} places these materials at the verge of reaching 
the long-dreamed room-temperature superconductivity. 
Electrical and x-ray diffraction measurements determined a weakly pressure-dependent \tc\ for \laH\ 
between 137 and 218 gigapascals in a structure with a face-centered cubic (fcc) arrangement of La atoms~\cite{Nature_LaH_Eremets_2019}. 
Here we show that quantum atomic fluctuations stabilize in all this pressure range 
a high-symmetry \fcc \ crystal structure consistent with experiments, which has a colossal electron-phonon 
coupling of $\lambda\sim3.5$. 
Even if {\it ab initio} classical calculations neglecting quantum atomic vibrations 
predict this structure to distort below 230\,GPa yielding a complex energy 
landscape with many local minima, the inclusion of quantum effects simplifies the energy landscape 
evidencing the \fcc \ as the true ground state. The agreement between the calculated and experimental \tc \ 
values further supports this phase as responsible for the 250\,K superconductivity. 
The relevance of quantum fluctuations in the energy landscape found here questions many of the crystal structure 
predictions made for hydrides within a classical approach that at the moment guide the experimental 
quest for room-temperature superconductivity~\cite{Review_Zurek-2018,Cini_Review_arxiv2019,Pickard_Review_ARCMP2019}.
Furthermore, quantum effects reveal crucial to sustain solids with extraordinary electron-phonon coupling that may otherwise be unstable~\cite{PRL_Ph-Softening_Allen_1972}.
\end{abstract}

\maketitle
%======================================================================== 150 words

%========================================================================
\begin{figure*}[ht!]
\begin{center}
\includegraphics[width=1.7\columnwidth]{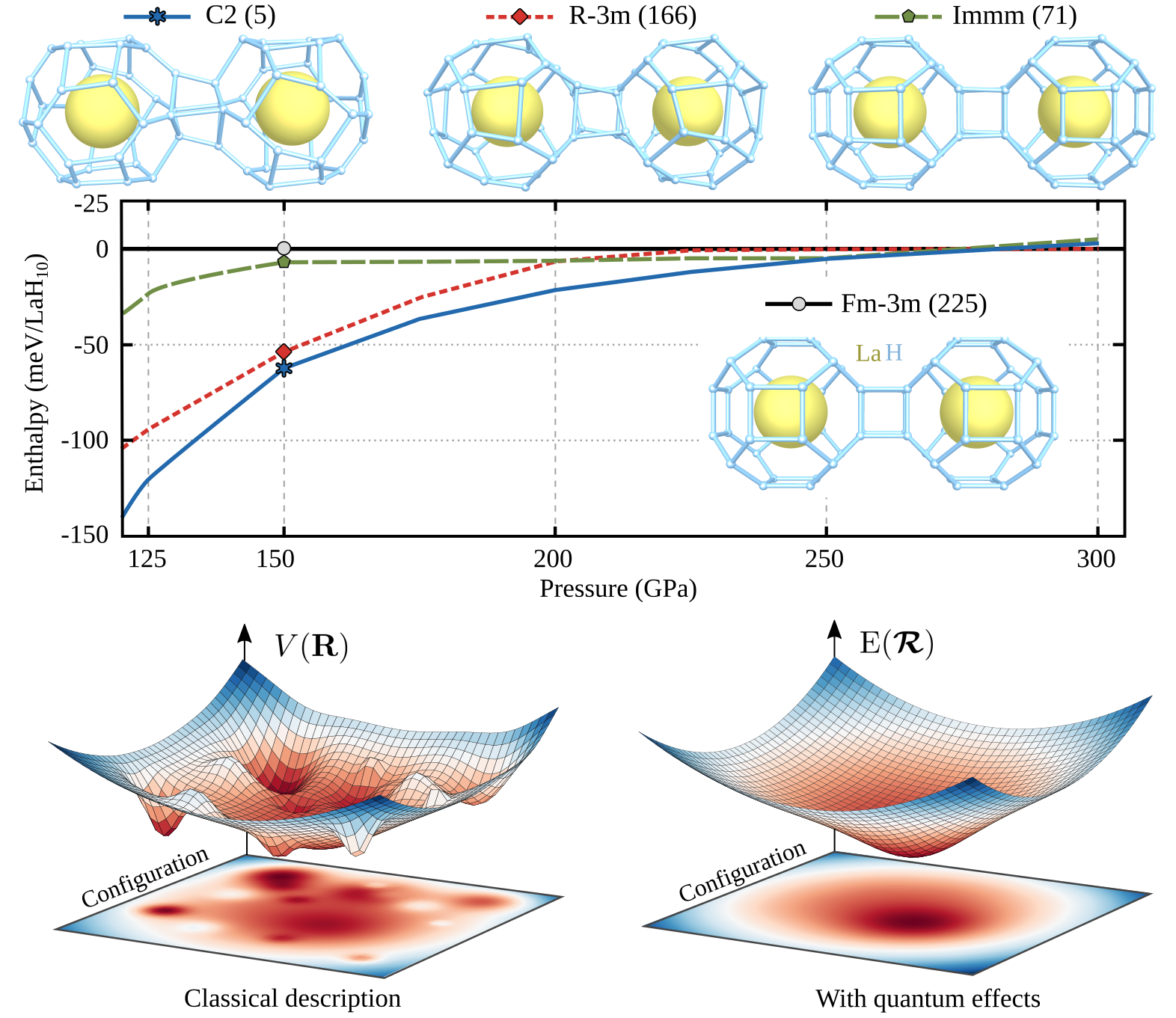}
\end{center}
\caption{~{\bf Quantum effects stabilize the symmetric \fcc\ phase of LaH$_{10}$.} 
Top panel: Enthalpy as function of pressure for different structures of LaH$_{10}$ calculated neglecting the zero-point energy. 
The pressure in the figure is calculated from $V(\mathbf{R})$, neglecting quantum effects on it. 
The crystal structure of the different phases found are shown. 
Bottom left: Sketch of a Born-Oppenheimer energy surface $V(\mathbf{R})$ exemplifying the presence of many local minima for many distorted structures. $\mathbf{R}$ represents the positions of atoms treated classically as simple points.
Bottom right: sketch of the configurational $E(\boldsymbol{\mathcal{R}})$ energy surface including quantum effects. $\boldsymbol{\mathcal{R}}$ represents the quantum centroid positions, which determine the center of the ionic wave functions, i.e., the average atomic positions.
All phases collapse to a single phase, the highly symmetric \fcc. }\label{fig:structure}
\end{figure*}

The potential of metallic hydrogen as a high-\tc\ superconductor~\cite{Ashcroft_PRL1968,Ginzburg_1969} 
was identified few years after the development of the Bardeen-Cooper-Schrieffer (BCS) theory, 
which explained superconductivity through the electron-phonon coupling mechanism. The main argument was that \tc\ can be maximized for light compounds due to their high vibrational frequencies. In view of the large pressures needed to metallize hydrogen~\cite{Diaseaal1579}, chemical precompression with heavier atoms~\cite{Gilman_1971_PRL_LiHF,Ashcroft_PRL2004} 
was suggested as a pathway to decrease the pressure needed to reach metallicity and, thus, superconductivity. 
These ideas have bloomed thanks to modern {\it ab initio} crystal structure prediction methods based on density-functional theory (DFT)~\cite{Cini_Review_arxiv2019,Ma_NatRevMaterials_2017,Review_Oganov-Pickard_2019}. Hundreds of hydrogen-rich compounds have been predicted to be thermodynamically stable at high pressures and, by calculating the electron-phonon interaction parameters, their \tc's have been estimated~\cite{Review_Zurek-2018,Cini_Review_arxiv2019}. The success of this symbiosis between DFT crystal structure predictions and \tc\ calculations is exemplified by the discovery of superconductivity in \sh\ at 200 K~\cite{DrozdovEremets_Nature2015,li2014metallization,Duan_SciRep2014}. 
The prospects for discovering warm hydrogen-based superconductors in the next years are thus high, 
in clear contrast with other high-\tc\ superconducting families such as cuprates or pnictides~\cite{Schilling_cuprate_nature1993,fesc:kamihara2006}, where the lack of a clear 
understanding of the superconducting mechanism hinders an {\it in silico} guided approach.   

DFT predictions in the La-H system proposed \laH\ to be thermodynamically stable against decomposition above 150\,GPa. 
A sodalite type-structure with space group \fcc\ and \tc$\sim$280\,K was suggested above $\sim$220 GPa 
(see Fig. \ref{fig:structure}), and a distorted version of it below with space group $C2/m$ and a rhombohedral La sublattice~\cite{PNAS_LaHx_2017_Hemley,geballe2018synthesis}. 
By laser heating a lanthanum sample in a hydrogen-rich atmosphere within a diamond anvil cell (DAC), a lanthanum superhydride was synthesized right after~\cite{geballe2018synthesis}. 
Based on the unit cell volume obtained by x-ray diffraction, the hydrogen to lanthanum ratio was estimated to be between 9 and 12. 
An fcc arrangement of the La atoms was determined above $\sim$160\,GPa, and a rhombohedral lattice below with $R$-$3m$ space group for the La sublattice. 
Due to the small x-ray cross section of hydrogen, experimentally it is not possible to resolve directly the H sublattice. 
Early this year, evidences of a superconducting transition at 260\,K  and 188\,GPa were reported 
in a lanthanum superhydride~\cite{Hemley-LaH10_PRL_2019}. These findings were confirmed and put in solid grounds 
few months later  by an independent group that measured a \tc\ of 250\,K from 137 to 218\,GPa in a structure with 
fcc arrangement of the La atoms and suggested a \laH\ stoichiometry~\cite{Nature_LaH_Eremets_2019}. 

\begin{figure*}[ht!]
\begin{center}
\includegraphics[width=1.0\textwidth]{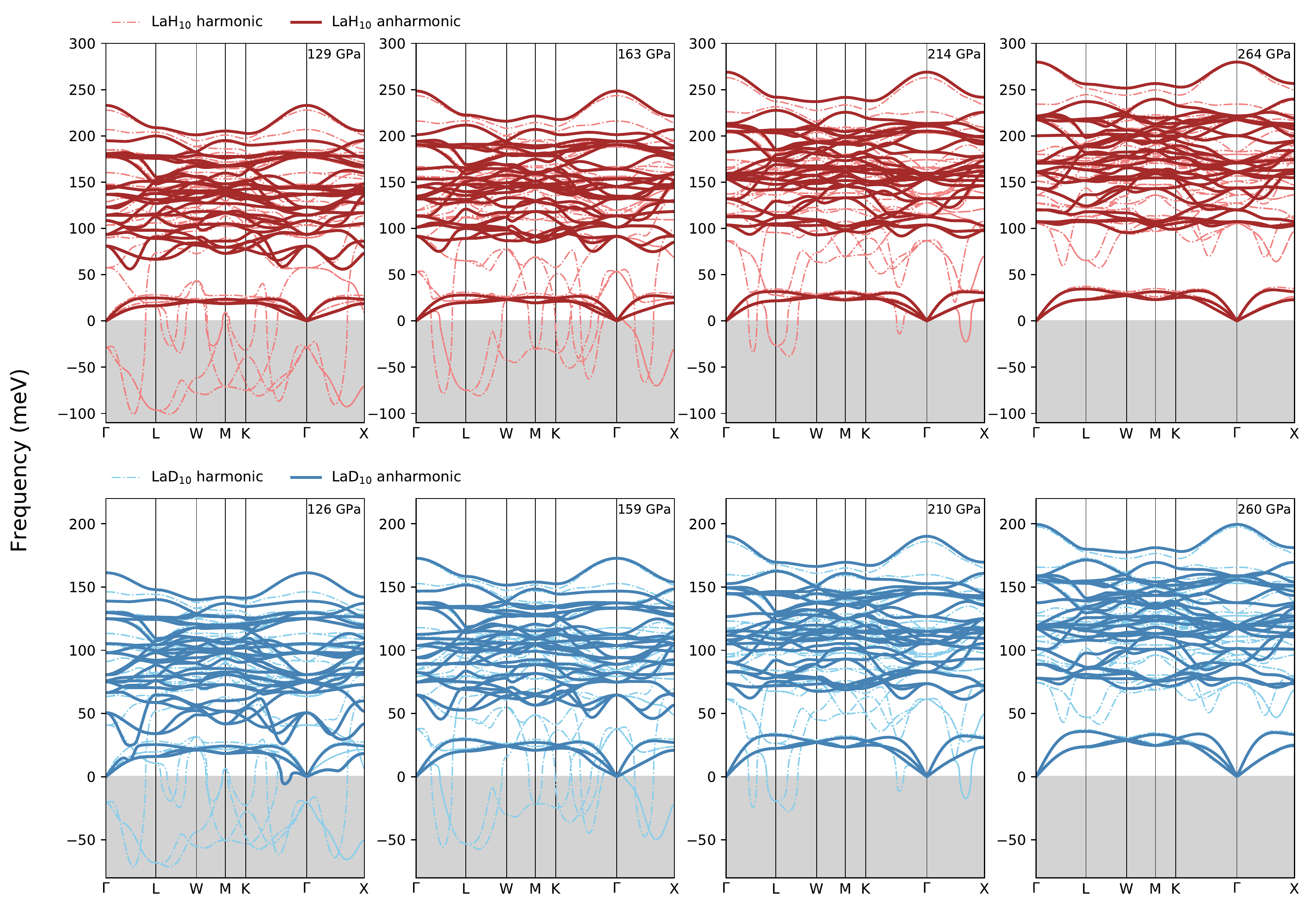}
\end{center}
\vspace{-0.5cm}
\caption{~{\bf Phonon band structure of \fcc\ LaH$_{10}$ at different pressures.}
The harmonic phonons show large instabilities in several regions of the Brillouin zone. 
Only at the high-pressure limit, e.g. above 220-250\,GPa dynamic (harmonic) stabilization is reached. 
The anharmonic phonons obtained from the Hessian of the quantum $E(\boldsymbol{\mathcal{R}})$ 
energy within the SSCHA are dynamically stable in the experimentally relevant range. 
The case of deuterium develops a instability at low pressures (126 GPa) consistent with experimental evidence. 
The pressure given corresponds to the calculated from $E(\boldsymbol{\mathcal{R}})$ that considers quantum effects. 
The grey area marks the region with imaginary phonon frequencies, which are depicted as negative frequencies.}\label{fig:Phonons}
\end{figure*}

Even if it is tempting to assign the record superconductivity to the \fcc\ phase predicted previously~\cite{Hemley-LaH10_PRL_2019,Nature_LaH_Eremets_2019}, there is a clear problem: 
the \fcc\ structure is predicted to be dynamically unstable in the whole pressure range where a 250\,K \tc\ was observed. 
This implies that this phase is not a minimum of the Born-Oppenheimer energy surface. 
Consequently, no \tc\ has been estimated for this phase in the experimental pressure range. 
Considering that quantum proton fluctuations symmetrize hydrogen bonds in the high-pressure X 
phase of ice~\cite{Benoit_H2O-symmetric_Nat1998} and in \sh~\cite{Nature_Errea_2016,H3S-anharm-Bianco}, 
this contradiction may signal a problem of the classical treatment of the atomic vibrations in the calculations. 
We show here how quantum atomic fluctuations completely reshape 
the energy landscape making the \fcc\ phase the true ground state and the responsible 
for the observed superconducting critical temperature. 

We start by calculating with DFT the lowest enthalpy structures of \laH\ as a 
function of pressure with state-of-the-art crystal structure prediction methods~\cite{Goedecker_mhm_2004,Amsler_mhm_2010}. 
The contribution associated with atomic fluctuations is not included, 
so that the energy just corresponds to the Born-Oppenheimer energy $V(\mathbf{R})$, where $\mathbf{R}$ represents the position of atoms treated classically as simple points. As shown in Figure~\ref{fig:structure}, different distorted phases of \laH\ are thermodynamically more stable than the \fcc\ phase. Above $\sim$250\,GPa all phases merge to the \fcc\ symmetric phase. 
These results are in agreement with previous calculations~\cite{PNAS_LaHx_2017_Hemley}, 
even if we identify other possible distorted structures 
with lower enthalpy such as 
the  $R$-$3m$, $C2$ and $P1$ (not shown) phases.  
These phases not only imply a distortion of the H atoms, also show a La sublattice without 
an fcc arrangement, and thus should be detectable by x-ray. 
The fact that many structures are predicted 
underlines that the classical $V(\mathbf{R})$ energy surface is 
of a multifunnel structure tractable to many different saddle and local minima, as sketched in Figure~\ref{fig:structure}. 

This picture completely changes when including the energy of quantum atomic fluctuations, 
the zero-point energy (ZPE). We calculate the ZPE within the stochastic self-consistent harmonic approximation (SSCHA)~\cite{SCHA-method_PRB_Errea_2014,Bianco_SSCHA_PRB2017,Monacelli_SSCHA_2018PRB}. 
The SSCHA is a variational method that calculates the $E(\boldsymbol{\mathcal{R}})$ energy of the system including 
atomic quantum fluctuations as a function of the {\it centroid} positions $\boldsymbol{\mathcal{R}}$, 
which determine the center of the ionic wave functions. The calculations are performed without 
approximating the $V(\mathbf{R})$ potential, keeping all its anharmonic terms. 
We perform a minimization of $E(\boldsymbol{\mathcal{R}})$ and determine the centroid positions at its minimum.
By calculating the stress tensor from $E(\boldsymbol{\mathcal{R}})$~\cite{Monacelli_SSCHA_2018PRB}, 
we relax the lattice parameters seeking for structures with isotropic stress conditions considering quantum effects. 
We start the quantum relaxation for both $R$-$3m$ and $C2$ phases with the lattice that yields a classical isotropic 
pressure of 150\,GPa and vanishing classical forces, i.e., calculated from $V(\mathbf{R})$. 
All quantum relaxations quickly evolve into the \fcc\ phase. 
This suggests that the quantum $E(\boldsymbol{\mathcal{R}})$ energy landscape is much simpler than the 
classical $V(\mathbf{R})$ as sketched in Figure~\ref{fig:structure}. 
And that the sodalite symmetric \fcc\ phase is the ground state for \laH\ in all the pressure range of interest. 
Quantum effects are colossal: reshaping the energy landscape and stabilizing structures by more than 60\,meV per \laH. 

Our results further confirm that the structure of \laH\ responsible for the 250\,K superconductivity is \fcc.  
This is completely consistent with the fcc arrangement of La atoms found experimentally~\cite{Nature_LaH_Eremets_2019}. 
However, Geballe et al.~\cite{geballe2018synthesis} observed a rhombohedral distortion below $\sim$160 GPa, 
with an $R$-$3m$ space group for the La sublattice and a rhombohedral angle of approximately 61.3\textdegree 
($c/a\sim2.38$ in the hexagonal representation). Our calculations show that this distortion is compatible with slight 
anisotropic stress conditions in the DAC. Indeed, performing a SSCHA minimization for our $R$-$3m$ 
phase but keeping the rhombohedral angle fixed at 62.3\textdegree (the value that yields an isotropic pressure of 150\,GPa 
at the classical level) the quantum stress tensor shows a 6\% anisotropy between the diagonal direction and the perpendicular plane. 
This suggests that anisotropic conditions inside the DAC can produce the $R$-$3m$ phase, 
while other experimental stress conditions could favor other crystal phases. 

The \fcc\ phonon spectra calculated in the harmonic approximation from the Hessian of $V(\mathbf{R})$ 
show clear phonon instabilities in a broad region of the Brillouin zone (see Figure~\ref{fig:Phonons}). 
These instabilities appear below $\sim$230\,GPa. This is consistent with the fact that below this pressure 
many possible atomic distortions lower the enthalpy of this phase. On the contrary, as shown in Figure~\ref{fig:Phonons}, 
when calculating the phonons from the Hessian of $E(\boldsymbol{\mathcal{R}})$~\cite{Bianco_SSCHA_PRB2017}, 
which effectively captures the full anharmonicity of $V(\mathbf{R})$, no instability is observed. This confirms again 
that the \fcc\ phase is a minimum in the quantum-energy landscape in the whole pressure range where 
a 250\,K \tc\ was observed. 
While the \fcc\ phase of \laH\ remains a minimum of $E(\boldsymbol{\mathcal{R}})$ as low as $\sim$ 129\,GPa, 
the case of \laD\ shows instabilities at 126\,GPa, 
implying that at this pressure the \fcc\ phase of LaD$_{10}$ distorts 
to a new phase (as suggested by Drozdov et al.~\cite{Nature_LaH_Eremets_2019}). 
Below this pressure we also predict that \laH\ composition is not 
longer thermodynamically stable 
and low-hydrogen compositions are likely to  occur.   

\begin{figure}
\begin{center}
\includegraphics[width=1\columnwidth]{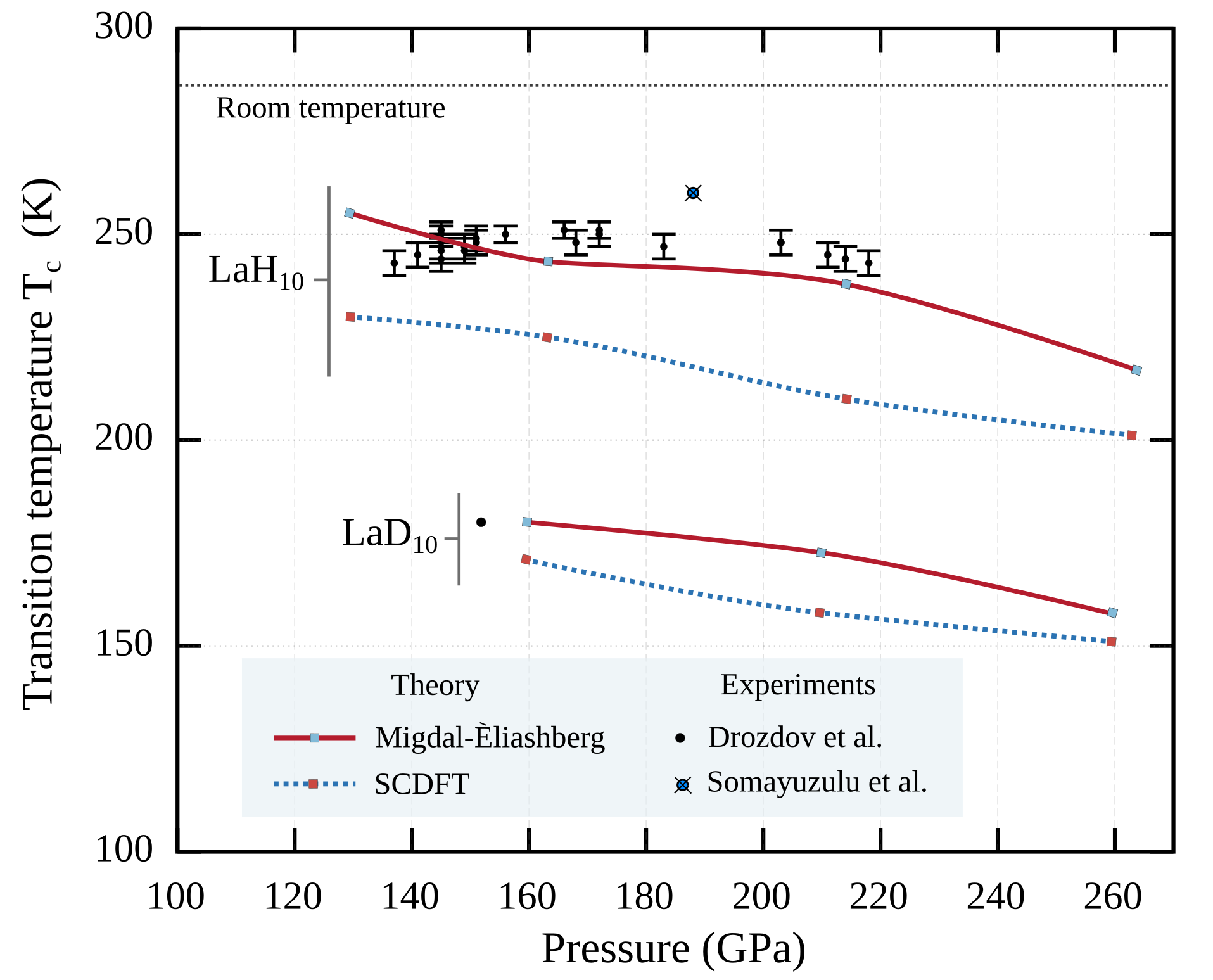}
\end{center}
\caption{~{\bf Summary of experimental and theoretical \tc\ values.} 
Superconducting critical temperatures calculated within anisotropic Migdal-\'Eliashberg equations and SCDFT. 
In both cases the anharmonic phonons obtained with the SSCHA are used. 
The results are compared with the experimental measurements 
by Somayuzulu et al.~\cite{Hemley-LaH10_PRL_2019} and Drozdov et al.~\cite{Nature_LaH_Eremets_2019}.}\label{fig:Tc}
\end{figure}

Flagrantly, the breakdown of the classical harmonic approximation for phonons makes impossible 
the estimation of \tc\ below $\sim$250 GPa in the \fcc\ phase and questions all previous calculations~\cite{PNAS_LaHx_2017_Hemley,Liu_MicroscopicMechanismLaH10_PRB2019}. 
Indeed, the anharmonic phonon renormalization remains huge also at 264\,GPa (see Figure~\ref{fig:Phonons}). 
On the contrary, with anharmonic phonons derived from the Hessian of $E(\boldsymbol{\mathcal{R}})$ 
we can readily calculate the electron-phonon interaction 
and the superconducting \tc\ in the experimental range of pressure (120--210\,GPa). 
The superconducting critical temperature is estimated fully {\it ab initio}  
--without any empirical parameter-- by solving Migdal-\'Eliashberg (ME) equations and applying SuperConducting DFT (SCDFT). 
As shown in Figure~\ref{fig:Tc}, the numerical solutions of ME equations with anisotropic energy gap are almost 
on top of the experimental values. SCDFT values systematically show a slightly lower \tc. 
Our reported values of \tc\ evidence the phonon-driven mechanism of superconductivity and confirm 
\laH\ in its \fcc\ structure as responsible for the highest-\tc\ up to date reported. 
Our calculations for \laD\ in the \fcc\ phase are also in agreement with the experimental point reported. 
Despite the large anharmonic effects at play, the isotope coefficient $\alpha=- \left[ \ln{\mathrm{T_{\text{c}}(LaD_{10})}} - \ln{\mathrm{T_{\text{c}}(LaH_{10})}} \right] / \ln 2$ is close to 0.5 (0.43 around 160 GPa), the expected value in BCS theory, 
and it is in agreement with the experimentally reported $\alpha=0.46$. 

We finally check \tc\ for the subtle rhombohedral distortion 
that could be induced by anisotropic stress conditions of pressure.  
Fixing the rhombohedral angle at 62.3\textdegree\ the obtained \tc\ 
for the $R$-$3m$ phase at 160\,GPa is a 9\% lower than for the \fcc. 
Thus, the observed weak pressure dependence of \tc\ is consistent with the absence of a rhombohedral distortion, 
as suggested by the x-ray data~\cite{Nature_LaH_Eremets_2019}. 
However, as argued above, undesired anisotropic stress conditions in the DAC can induce phase transitions. 
We thus believe that other experimental \tc\ measurements with lower values but around 200\,K 
correspond to distorted structures induced by anisotropic conditions of pressure. 
In fact, we can safely rule out that compositions such as LaH$_{11}$, 
proposed to yield a high critical temperature~\cite{Nature_LaH_Eremets_2019}, 
is responsible for any sizable \tc\ (see Extended Data).   

In summary, this work demonstrates how quantum effects are of capital importance
in determining the ground state structures of superconducting hydrides,
challenging all current predictions and evidencing flaws in standard theoretical methods.
It also illustrates that quantum fluctuations are indispensable to sustain crystals 
with huge $\lambda$'s ($\lambda$ reaches a record value of 3.6 at 129\,GPa for \laH) 
that be otherwise destabilized by the colossal electron-phonon interaction 
to distorted (low symmetry) structures reducing the electronic density 
of states at the Fermi level (see Extended Data)~\cite{PRL_Ph-Softening_Allen_1972}.  
This is relevant since large $\lambda$ is required to guarantee high-\tc~\cite{Cini_Review_arxiv2019,Pickard_Review_ARCMP2019}, not simply light atomic masses.

%========ANTO=KEEP=THIS=ALL===================
%{\bf In summary, we have shown that quantum effects in the lattice dynamics  are essential to establish the true 
% ground state structure of superconducting hydrides,
%challenging all current predictions and evidencing flaws in standard theoretical methods.
%It also illustrates that quantum fluctuations are indispensable to sustain crystals 
%with huge $\lambda$'s (reaching a record value of 3.6 at 129\,GPa for \laH). 
%This electron-phonon coupling strength, key to reach high \tc~\cite{Cini_Review_arxiv2019,Pickard_Review_ARCMP2019}, 
%in the harmonic approximation is limited by the lattice stability that could stabilize to a low %symmetry one with lower Fermi density of states, coupling and \tc. 
%The quantum lattice response prevents the instability allowing for much stronger electron phonon couplings and  broadening chances to room-temperature superconductivity. }
%

{\bf Acknowledgements.} 
This research was supported by the European Research Council (ERC) 
under the European Unions Horizon 2020 research and innovation programme (grant agreement No. 802533); 
the Spanish Ministry of Economy and Competitiveness (FIS2016-76617-P);
NCCR MARVEL funded by the Swiss National Science Foundation; and 
grant-in-Aid for Scientific Research (No.16H06345, 18K03442 and No. 19H05825) 
from the Ministry of Education, Culture, Sports, Science and Technology, Japan.
Computational resources were provided by 
the Barcelona Superconducting Center (project FI-2019-1-0031) and 
the Swiss National Supercomputing Center (CSCS).

{\bf Author contributions.} 
The project was conceived by I.E. and J.A.F.-L. 
The SSCHA was developed by 
I.E., L.M., R.B., M.C. and F.M. In particular, R.B. developed the method to compute
the quantum energy Hessian and the anharmonic phonon dispersions, and L.M. developed the method to perform a quantum relaxation of the lattice parameters. 
I.E. and F.B. performed the SSCHA calculations. A.S. T.K., T.T., R.A. and J.A.F.-L. 
conducted studies on structure prediction and superconductivity. 
All authors contributed to the editing of the manuscript.

{\bf Competing interests.} The authors declare no competing interests. \\

% \clearpage

% \newpage

%----------------------------------
                     {\sc Methods}\\
%----------------------------------        

{\bf Calculation details.} 
First-principles calculations were performed within DFT and the generalized gradient approximation (GGA) as parametrized by Perdew, Burke, and Ernzerhof (PBE)~\cite{PBE_PRL1996}. Harmonic phonon frequencies were calculated within density functional perturbation theory (DFPT)~\cite{DFPT_S.Baroni} making use of the {\sc Quantum ESPRESSO} code~\cite{QE-2009,Quantumespresso_2017}. The SSCHA~\cite{Errea_PdH_PRL2013,SCHA-method_PRB_Errea_2014,Bianco_SSCHA_PRB2017,Monacelli_SSCHA_2018PRB} minimization requires the calculation of energies, forces and stress tensors in supercells. These were calculated as well within DFT at the PBE level with {\sc Quantum ESPRESSO}. For the final SSCHA populations, 1000 configurations were used to reduce the stochastic noise. In all these calculations we used ultrasoft pseudopotentials including 11 electrons for the La atoms, a plane-wave cut-off energy of 50\,Ry for the kinetic energy and 500\,Ry for the charge density.

In the harmonic phonon calculations for the \fcc\ and $R$-$3m$ phases, we used the primitive and rhombohedral lattices, respectively, with one \laH\ formula unit in the unit cell. A 20$\times$20$\times$20 Monkhorst-Pack shifted electron-momentum grid was used for these calculations with a Methfessel-Paxton smearing of 0.02\,Ry. The DFT calculations performed for the SSCHA on supercells were performed on a coarser electron-momentum grid, which would correspond to a 12$\times$12$\times$12 grid in the unit cell. We explicitly verified that this coarser mesh yields a fully converged SSCHA gradient with respect to the electron-momentum grid, thus, not affecting the SSCHA minimization. The DFT supercell calculations for the SSCHA minimization on the $C2$ phase were performed keeping the same $\mathbf{ k}$-point density.

All phonon frequencies for $\mathbf{q}$-points not commensurate with the supercell used in the SSCHA minimization were obtained by directly Fourier interpolating the real space force constants obtained in this supercell, which are calculated form the Hessian of $E(\boldsymbol{\mathcal{R}})$. For the \fcc\ phase the SSCHA calculation was performed both on a 2$\times$2$\times$2 and 3$\times$3$\times$3 supercell containing, respectively, 88 and 297 atoms. The phonon spectra shown in Figure \ref{fig:Phonons} for the \fcc\ phase are obtained by Fourier interpolating directly the SSCHA energy Hessian force constants obtained in a 3$\times$3$\times$3 supercell. In Extended Data Figure \ref{fig:LaH10_supercell} we show that the phonon spectrum obtained interpolating directly the force constants in a 2$\times$2$\times$2 supercell yields similar results, indicating that the energy Hessian force constants are short-range and can be Fourier interpolated. Indeed, the \tc\ calculated with the 2$\times$2$\times$2 and 3$\times$3$\times$3 force constants for interpolating phonons only differs in approximately 3 K. Upon this, the SSCHA quantum structural relaxations in the $R$-$3m$ and $C2$ phases were performed in supercells with 88 atoms. 

As shown in Ref.~\cite{Bianco_SSCHA_PRB2017}, the Hessian of $E(\boldsymbol{\mathcal{R}})$ is
\begin{equation}
   \frac{\partial^2 E(\boldsymbol{\mathcal{R}})}{\partial \boldsymbol{\mathcal{R}} \partial \boldsymbol{\mathcal{R}}} = \boldsymbol{\Phi} + \overset{(3)}{\boldsymbol{\Phi}} \boldsymbol{\Lambda} \left[ \mathbf{1} - \overset{(4)}{\boldsymbol{\Phi}} \boldsymbol{\Lambda}  \right]^{-1} \overset{(3)}{\boldsymbol{\Phi}}. \label{eq:Hessian}
\end{equation} 
Bold notation represents matrices and tensors in compact notation. In Eq.~\eqref{eq:Hessian},  $\boldsymbol{\Phi}$ are the variational force constants of the SSCHA minimization, $\overset{(n)}{\boldsymbol{\Phi}}$ the quantum statistical averages taken with the SSCHA density matrix of the $n$-th order derivatives of $V(\mathbf{R})$, and $\boldsymbol{\Lambda}$ a tensor that depends on the temperature and  $\boldsymbol{\Phi}$. $\mathbf{1}$ is the identity matrix.  As we show in Extended Figure \ref{fig:LaH10_d4v}, setting $\overset{(4)}{\boldsymbol{\Phi}}=0 $ has a negligible effect on the phonons obtained from the Hessian defined in Eq. \eqref{eq:Hessian}. Therefore, $\overset{(4)}{\boldsymbol{\Phi}}$ is neglected throughout, and all superconductivity calculations in the \fcc\ and $R$-$3m$ phases are performed making use of the phonon frequencies and polarization vectors obtained from the Hessian of $E(\boldsymbol{\mathcal{R}})$ with $\overset{(4)}{\boldsymbol{\Phi}}=0 $. We also estimated \tc\ with the phonon frequencies and polarization vectors obtained instead from $\mathbf{\Phi}$, resulting in a critical temperature 12\,K lower within Allen-Dynes modified McMillan equation. This difference is small and within the uncertainty of the \tc\ calculation between SCDFT and anisotropic Migdal-\'Eliashberg calculations (see Figure \ref{fig:Tc} and below).

The \'Eliashberg spectral function, which we used for the \tc\ calculations, is defined as 
\begin{eqnarray}
 \alpha^2 F(\omega) & = & \frac{1}{N_{E_F}} \sum \limits_{n\mathbf{k}, m\mathbf{q},\nu} |g^{\nu}_{n\mathbf{k},m\mathbf{k}+\mathbf{q}}|^2 \delta(\epsilon_{n\mathbf{k}}-E_F)  \nonumber \\ & \times & \delta(\epsilon_{m\mathbf{k}+\mathbf{q}}-E_F) \delta(\omega-\omega_{\mathbf{q}\nu}), \label{eq:a2F}
\end{eqnarray}
where $N_{E_F}$ is the electronic density of states (DOS) at the Fermi energy ($E_F$), $n$ and $m$ are band indices, $\mathbf{k}$ is a crystal momentum, $\epsilon_{n\mathbf{k}}$ is a band energy, $\omega_{\mathbf{q}\nu}$ is the phonon frequency of mode $\nu$ at wavevector $\mathbf{q}$, and $g^{\nu}_{n\mathbf{k},m\mathbf{k}+\mathbf{q}}$ is the electron-phonon matrix element between a state $n\mathbf{k}$ and $m\mathbf{k}+\mathbf{q}$. We calculated $\alpha^2 F(\omega)$ combining the SSCHA  phonon frequencies and polarization vectors obtained from the Hessian of $E(\boldsymbol{\mathcal{R}})$ with the electron-phonon matrix elements calculated with DFPT. For the \fcc\ and $R$-$3m$ phases, the electron-phonon matrix elements were calculated in a 6$\times$6$\times$6 $\mathbf{q}$ point grid and a 40$\times$40$\times$40 $\mathbf{k}$ point grid. These were combined with the SSCHA phonons and polarization vectors obtained by Fourier interpolation to the 6$\times$6$\times$6 $\mathbf{q}$ point grid from the real space force constants coming from the Hessian of $E(\boldsymbol{\mathcal{R}})$ in a 3$\times$3$\times$3 supercell for the \fcc\ and in a 2$\times$2$\times$2 supercell for the $R$-$3m$. The Dirac deltas on the band energies are estimated by substituting them with a Gaussian of 0.004\,Ry width. The calculated  $\alpha^2 F(\omega)$ functions for the \fcc\ phase are shown in Extended Figures \ref{fig:LaH10_fcc_a2f} and \ref{fig:LaD10_fcc_a2f}, while in Extended Figure \ref{fig:R-3m_Ph_LaH10} we show the results for the $R$-$3m$ phase.

{\bf Crystal phase diagram exploration.}
To sample the enthalpy landscape of LaH$_{10}$ we employed the minima hopping method 
(MHM)~\cite{Goedecker_mhm_2004,Amsler_mhm_2010}, which has been successfully used for 
global geometry optimization in a large variety of applications 
including superconducting materials such as H$_3$S, PH$_3$, and disilane  
at high pressure~\cite{flores-sanna_HSe_2016,flores-sanna_PH3_2016}. 
This composition was thoroughly explored with 1, 2, 3 and 4 formula units simulation cells. 
Variable composition simulations were also performed for other La-H compositions. 
Energy, atomic forces and stresses were evaluated at the DFT level with the GGA-PBE parametrization 
to the exchange-correlation functional. A plane wave basis-set with a high cutoff energy of 
900\,eV was used to expand the wave-function together with the projector augmented wave (PAW) 
method as implemented in the Vienna Ab Initio Simulation Package~{\sc vasp}~\cite{VASP_Kresse}. 
Geometry relaxations were performed with tight convergence criteria 
such that the forces on the atoms were less than 2\,meV/\AA\ and the stresses were less than 0.1~eV/\AA$^3$. 
Extended Data Figure \ref{fig:Convex-hull} shows our calculated convex hull of enthalpy formation 
without considering the zero-point energy at 100, 150 and 200\,GPa. 
Interestingly, there are many stable compositions in the convex hull. 
\laH\ becomes enthalpically stable at $\sim$125\,GPa and remains in the convex well above 300\,GPa. 
Below 150\,GPa, $R$-$3m$ and $C2$ phases (\laH) show unstable harmonic phonon at $\Gamma$, 
becoming saddle points of $V(\mathbf{R})$. 
However, harmonically one can find $P1$ stable structures (decreasing symmetry) by 
following the instability pattern (softening direction, i.e. along eigenvector polarization). 
$P1$ structures are degenerate in enthalpy within less 3\,meV/\laH\ with respect $C2$.
Hence, we used the $C2$ as a representative of highly distorted structures for our study. \\ 

{\bf Superconductivity calculations in the \fcc\ phase.}
Superconductivity calculations were performed within two different approaches that represent the state-of-the-art of {\it ab initio} superconductivity: Density functional theory for Superconductors (SCDFT) and the \'Eliashberg equations with full Coulomb interaction. 

SCDFT is an extension to DFT for a superconducting ground state~\cite{OGK_SCDFT_PRL1988,Lueders_SCDFT_PRB2005}.
By assuming that the ${n{\bf k}}$ anisotropy in the electron-phonon coupling is negligible (see Ref.~\onlinecite{Lueders_SCDFT_PRB2005} for further details), the critical temperature is computed by solving an (isotropic) equation for the Kohn-Sham gap:
\begin{equation}\label{eq:gapSCDFT}
\Delta_s\left(\epsilon\right)=\mathcal{Z}\left(\epsilon\right)\Delta_s\left(\epsilon\right)
-\int d\epsilon' 
\mathcal{K}\left(\epsilon,\epsilon'\right)\frac{\tanh\left[\frac{\beta E\left(\epsilon'\right)}{2}\right]}{2E\left(\epsilon'\right)}\Delta_s\left(\epsilon'\right),
\end{equation}
where $\epsilon$ is the electron energy and $\beta$ the inverse temperature. 
The kernels $\mathcal{K}$ and $\mathcal{Z}$ come from the exchange correlation functional of the theory~\cite{flores-Sanna_honeycombs_2015,SCDFTWO3_PRM2019,Lueders_SCDFT_PRB2005,Marques_SCDFT_PRB2005,Linscheid_localOP_PRL2015,Massidda_SUST_CoulombSCDFT_2009} and depend on the properties of the pairing interactions: 
electron-phonon coupling and screened electron-electron repulsion. 
Eq.~\ref{eq:gapSCDFT} allows us to calculate \tc\ completely {\it ab initio}, without introducing an
empirical $\mu^*$ parameter (Coulomb pseudopotential). 
Dynamic effects on the Coulomb interaction (plasmon) were also tested and did not show any significant effect. 
In its isotropic form, the screened Coulomb interaction in SCDFT is accounted for by a function $\mu(\epsilon,\epsilon')$, 
which is given by the average~\cite{Sanna_Eliashberg_JPSJ2018} 
RPA Coulomb matrix element on the iso-energy surfaces $\epsilon$ and $\epsilon'$ times the DOS at $\epsilon'$ ($N(\epsilon')$):
\begin{equation} 
\mu(\epsilon,\epsilon')=\sum_{n,m}\iint d^3(k k') V^{RPA}_{n{\bf k},m{\bf k}'}\frac{\delta({\epsilon-\epsilon_{n{\bf k}}})}{N(\epsilon_{n{\bf k}})} \delta({\epsilon'-\epsilon_{m{\bf k}'}}).
\end{equation}
The full energy dependence of the DOS is accounted in the calculations, while the electron-phonon coupling is described by the $\alpha^2 F(\omega)$ of Eq.~\eqref{eq:a2F}.

The second approach we use to simulate the superconducting state is the anisotropic \'Eliashberg approach~\cite{PRB_Sano_Van-Hove_H3S_2016}. 
Here we include, together with the energy dependence of the electron DOS, the anisotropy of the electron-phonon coupling. 
The Green's function form of the \'Eliashberg equation~\cite{PRB_Sano_Van-Hove_H3S_2016} we solve is given as
\begin{align}
    \Sigma_{n \mathbf{k}}(i \omega_i) &= 
    -\frac{1}{N \beta} \sum_{\mu,\mathbf{q},m} V_{mn}^{\rm ph}(\mathbf{q}, i\omega_\mu) G_{m \mathbf{k}+\mathbf{q}}(i\omega_\mu + i\omega_i),\label{eq:eliashberg_sigma}\\
    \Delta_{n \mathbf{k}}(i \omega_i) &= 
    -\frac{1}{N \beta} \sum_{\mu,\mathbf{q},m} \{ V_{mn}^{\rm ph}(\mathbf{q}, i\omega_\mu) + V_{mn}^{\rm C}(\mathbf{q}, i\omega_\mu) \} \nonumber\\
    &\times | G_{m \mathbf{k}+\mathbf{q}}(i\omega_\mu + i\omega_i) |^2 \Delta_{m \mathbf{k}+\mathbf{q}}(i\omega_\mu+i\omega_i).\label{eq:eliashberg_delta}
\end{align}
Here, $\Sigma_{n \mathbf{k}}(i\omega_i)$ and $\Delta_{n \mathbf{k}}(i \omega_i)$ are the normal and anomalous self energy, and $V_{mn}^{\rm ph}(\mathbf{q}, i\omega_\mu)$ and $V_{mn}^{\rm C}(\mathbf{q}, i\omega_\mu)$ are the $\mathbf{k}$-averaged phonon-mediated interaction and Coulomb interaction, respectively.
The explicit form of $V_{mn}^{\rm ph}(\mathbf{q}, i\omega_\mu)$ is given as 
\begin{align}
V_{mn}^{\rm ph}(\mathbf{q}, i\omega_\mu) = \sum_\nu | g_{nm}^\nu(\mathbf{q}) |^2 D_{\nu}(\mathbf{q},i\omega_\mu),
\end{align}
where $|g_{nm}^\nu(\mathbf{q})|^2$ is a ${\mathbf{k}}$-averaged electron-phonon matrix element,
\begin{align}
    |g_{nm}^\nu(\mathbf{q})|^2 = \frac{\sum_{\mathbf{k}} |g^{\nu}_{n\mathbf{k},m\mathbf{k}+\mathbf{q}}|^2 \delta(\epsilon_{n\mathbf{k}}-E_F) \delta(\epsilon_{m\mathbf{k}+\mathbf{q}}-E_F)}
    {\sum_{\mathbf{k}} \delta(\epsilon_{n\mathbf{k}}-E_F) \delta(\epsilon_{m\mathbf{k}+\mathbf{q}}-E_F)},
\end{align}
and $D_{\nu}(\mathbf{q}, i\omega_\mu)$ is a free-phonon Green's function, $D_{\nu}(\mathbf{q}, i\omega_\mu) = -2 \omega_{\mathbf{q} \nu}/(\omega_\mu^2 + \omega^2_{\mathbf{q} \nu})$.
The electron-phonon matrix elements are calculated through a DFPT calculation with 6$\times$6$\times$6 $\mathbf{q}$ point grid, and are combined with the phonon frequencies and polarization vectors obtained by directly Fourier interpolating to this grid the force constants coming from the $E(\boldsymbol{\mathcal{R}})$ Hessian in the 3$\times$3$\times$3 supercell. 
For the Coulomb interaction, $V_{mn}^{\rm C}(\mathbf{q}, i\omega_\mu)$ is approximated by ${\mathbf{k}}$-averaged static Coulomb interaction within the random phase approximation, $\frac{1}{N_k} \sum_{\mathbf{k}} V^{\rm RPA}_{m \mathbf{k}, n \mathbf{k+q}}(i\omega_\mu = 0)$.
Using Eq.~\eqref{eq:eliashberg_sigma}, the Dyson equation was solved self-consistently and then Eq.~\eqref{eq:eliashberg_delta} was solved to estimate $T_c$ with 36$\times$ 36 $\times$ 36 $\mathbf{k}$ point grid and 512 Matsubara frequencies.

In Extended Data Table~\ref{Tab:Master} we summarize all calculated \tc's within anisotropic ME and isotropic SCDFT. We also include the values obtained with McMillan equation and Allen-Dynes modified McMillan equation ($\mu^*$=0.1). The calculated electron-phonon coupling constant, $\lambda=2\int_0^{\infty}d\omega \alpha^2F(\omega)/\omega$, 
and the logarithmic frequency average, 
\omlog$=\exp \left( \frac{2}{\lambda} \int_0^{\infty}d\omega \frac{\alpha^2F(\omega)}{\omega}\log\omega  \right)$, 
are also included in the table.

{\bf Quantum structural relaxations in the $R$-$3m$ and $C2$ phases.}
In Extended Figure~\ref{fig:R-3m_IntRel_Pressure_LaH10} we show the evolution of the pressure calculated along the different Cartesian directions for the $R$-$3m$ throughout the SSCHA minimization but keeping the rhombohedral angle fixed at 62.3\textdegree. Thus, the centroid positions $\boldsymbol{\mathcal{R}}$ are optimized only considering the internal degrees of freedom of the $R$-$3m$ phase. Even if at the classical level the stress is isotropic (within a 0.5\%), after the SSCHA quantum relaxation an anisotropic stress of a 6\%  is created between the $z$ and $x-y$ directions. The phonons obtained at the end of the minimization are shown in Extended Figure~\ref{fig:R-3m_Ph_LaH10}. Secondly, in Extended Figure~\ref{fig:R-3m_CellRelax_LaH10}, we show that starting from the result of this minimization but now relaxing also the lattice, the $R$-$3m$ phase evolves into the \fcc\ phase. It is clear how the pressure calculated with quantum effects  becomes isotropic when the rhombohedral angle becomes 60\textdegree, the angle corresponding to a fcc lattice in a rhombohedral description. Also it is evident that the Wyckoff positions of the $R$-$3m$ phase evolve clearly into the \fcc\ Wyckoff positions, which are summarized in Extended Data Table~\ref{tab:LaH10_Wyckoff_tab}. 

In Extended Data Figure~\ref{fig:C2_Relax_Pressure_LaH10} we show the evolution of the diagonal components of the pressure along the three different Cartesian directions for the monoclininc $C2$ when the lattice structure is relaxed with the SSCHA. The starting point is obtained by first performing a SSCHA relaxation of only internal atomic coordinates keeping the lattice parameters that yield an isotropic stress of 150\,GPa. It is clear that quantum effects create an anisotropic stress if the lattice parameters are not modified. When the quantum relaxation of the lattice is performed, the lattice parameters are modified and an isotropic stress is recovered.

Extended Data Figure~\ref{fig:Ini-Fin_structures} shows the structures of the $R$-$3m$ and $C2$ phases obtained classically and after the quantum SSCHA relaxation. After the quantum relaxation, the symmetry of both structures is recognized as \fcc\ with a tolerance of 0.001\,\AA\ for lattice vectors and 0.005\,\AA\ for ionic positions, consistent with the stochastic accuracy of the SSCHA. 
In the same Figure~\ref{fig:Ini-Fin_structures}, 
the electronic density of states (DOS) as a function of pressure is plotted. 
Highly symmetric motif ($Fm$-$3m$) maximizes $N_{E_{F}}$, while in distorted structures 
($R$-$3m$ and $C2$) the occupation at the Fermi level 
is reduced by more than 20~\%. This underlines that the classical distortions would lower $N_{E_{F}}$, reducing $\lambda$, as expected in a system destabilized by the electron-phonon interaction.

{\bf Transition temperatures from other La-H compositions.}
Different compositions on the La-H phase diagram have been reported as thermodynamically stable. 
Presumably, the stabilization of these compositions and the measurement of different \tc's 
(see Drozdov et al.~\cite{Nature_LaH_Eremets_2019}) demonstrate that 
other stoichiometries are responsible for these measured \tc's.  
Notably, these \tc's appear substantially lower, for instance the values decrease from 250\,K, to 215\,K, 110\,K 
and to 70\,K. Experimentally there is not a clear correlation 
between sample preparation, \tc\ and pressure. 
In sample preparation of Drozdov et al. pressures can vary from 100 to 200\,GPa 
(gradient inside the diamond anvil cell) and it was proposed that other 
stoichiometries (low-hydrogen content) are responsible for systematically lower \tc's. 

Conversely, in a later publication the same authors suggested other hydrogen rich system  
that is enthalpically competitive (LaH$_{11}$)  and possibly responsible for other high-\tc\ phases. 
In order to explore this possibility, we did consider structure prediction runs 
with this stoichiometry and found crystalline structures that were previously 
reported in Ref.~\cite{Clathrate_REHX_PRL_2017}. 
Extended Data Figure~\ref{fig:LaH11} shows the structural motif and the 
corresponding phonons and $\alpha^2 F(\omega)$ spectral function. 
We can rule out the possibility that high-\tc, as measured in different samples,  
arises from LaH$_{11}$ in its $P4/nmm$ (129) structure 
(lowest enthalpy structure for this composition at relevant experimental pressures).  
As seen in Extended Data Figure~\ref{fig:LaH11}, this phase has a strong molecular crystal character, 
composed of H$_2$ units weakly interacting with La-lattice. 
This phase is indeed a poor metal with low occupation of electrons at the Fermi level 
due to its molecular character and cannot explain 70\,K, or higher values of \tc. 
Our estimated \tc\ with Allen-Dynes formula, harmonic phonons and using a $\mu^*$=0.1 is 7\,K at 100\,GPa. 

% \putbib[main]
% \end{bibunit}

\setcounter{figure}{0} 
\renewcommand{\figurename}{Extended Data Figure}
\renewcommand{\tablename}{Extended Data Table}

% \newpage

%-------------------------------------
% Extended data figures
%-------------------------------------

\begin{figure*}[ht!]
\begin{center}
\includegraphics[width=1.0\textwidth]{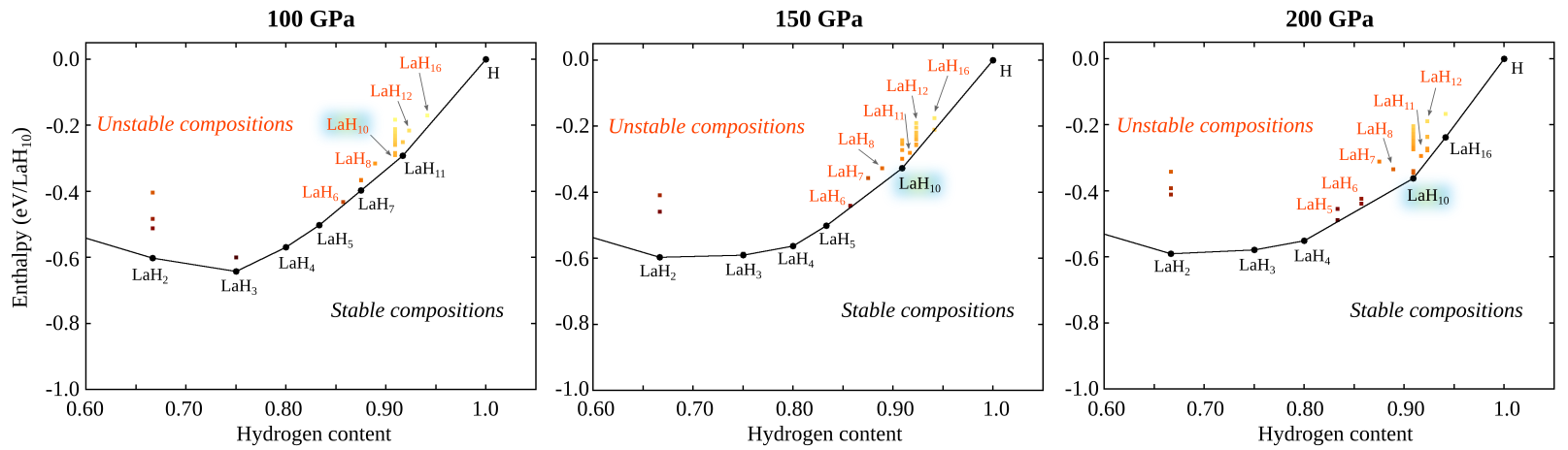}
\end{center}
\caption{~{\bf Convex hull of enthalpy formation.}
It is noticeable that at low pressure (left panel, 100\,GPa) 
the composition of LaH$_{10}$ is not stable and 
only develops as stable point in the convex hull above $\sim$125\,GPa.}\label{fig:Convex-hull}
\end{figure*}

%\newpage

\begin{figure*}
\begin{center}
\includegraphics[width=0.87\textwidth]{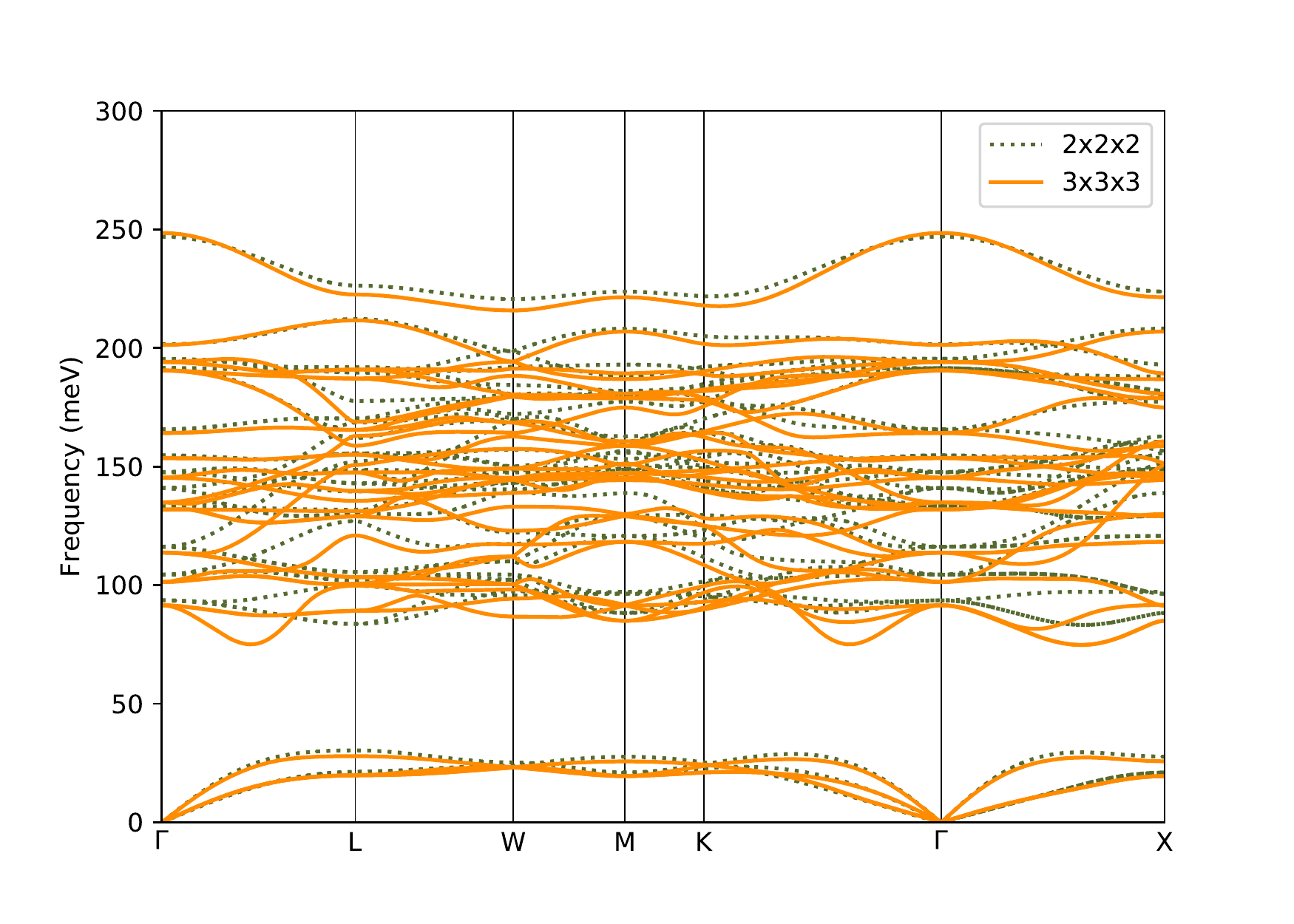}
\end{center}
\caption{~{\bf SSCHA phonons supercell convergence for \laH\ at 163 GPa.} 
The phonon spectra shown are calculated by directly Fourier interpolating the force constants obtained from the Hessian of $E(\boldsymbol{\mathcal{R}})$ in a real space $2\times2\times2$ and a $3\times3\times3$ supercell. The similarity of both phonon spectrum obtained by Fourier interpolation indicates that these SSCHA force constants are short-ranged and can be Fourier interpolated.}\label{fig:LaH10_supercell}
\end{figure*}

%\newpage

\begin{figure*}[ht!]
\begin{center}
\includegraphics[width=1.0\textwidth]{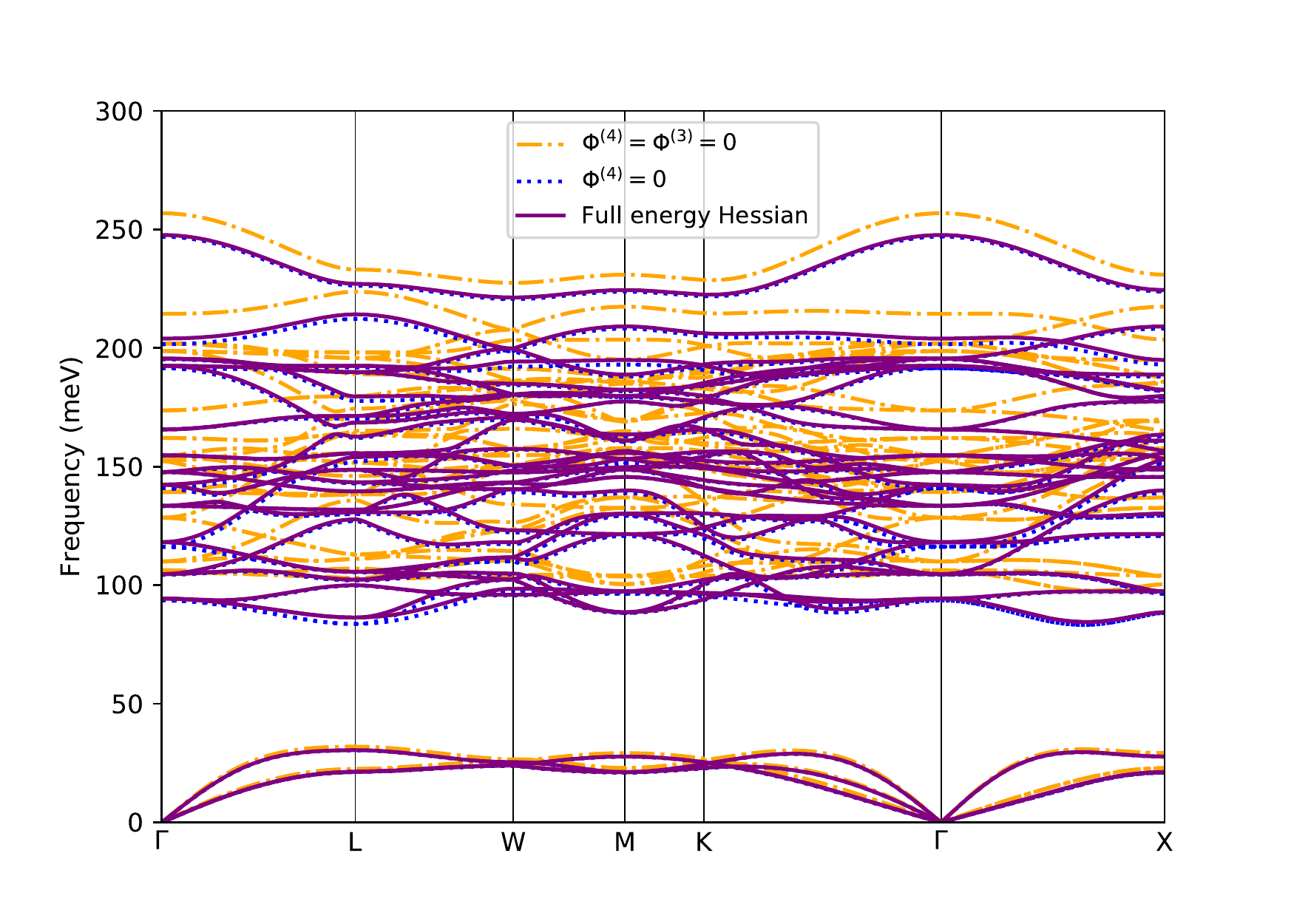}
\end{center}
\caption{~{\bf Different anharmonic phonons calculated  for \laH\ at 163 GPa.} 
Phonon spectra obtained from the SSCHA energy Hessian of Eq.~\eqref{eq:Hessian} making different level of approximations. The purple solid line is the phonon spectrum calculated with the full energy Hessian, without any approximation. In the blue dotted spectrum we set $\overset{(4)}{\boldsymbol{\Phi}}=0$ in the equation. In the orange dash-dotted line we set $\overset{(3)}{\boldsymbol{\Phi}}=\overset{(4)}{\boldsymbol{\Phi}}=0$, so that the phonon spectra corresponds to the one coming directly from the SSCHA variational force constants $\mathbf{\Phi}$. The results clearly show that while the effect of $\overset{(3)}{\boldsymbol{\Phi}}$ is important, setting $\overset{(4)}{\boldsymbol{\Phi}}=0$ is perfectly safe. All these phonon spectra in the figures are obtained by Fourier interpolating directly the real space anharmonic force constants in a $2\times2\times2$ supercell.
}\label{fig:LaH10_d4v}
\end{figure*}

\newpage

\begin{figure*}[ht!]
\begin{center}
\includegraphics[width=1.0\textwidth]{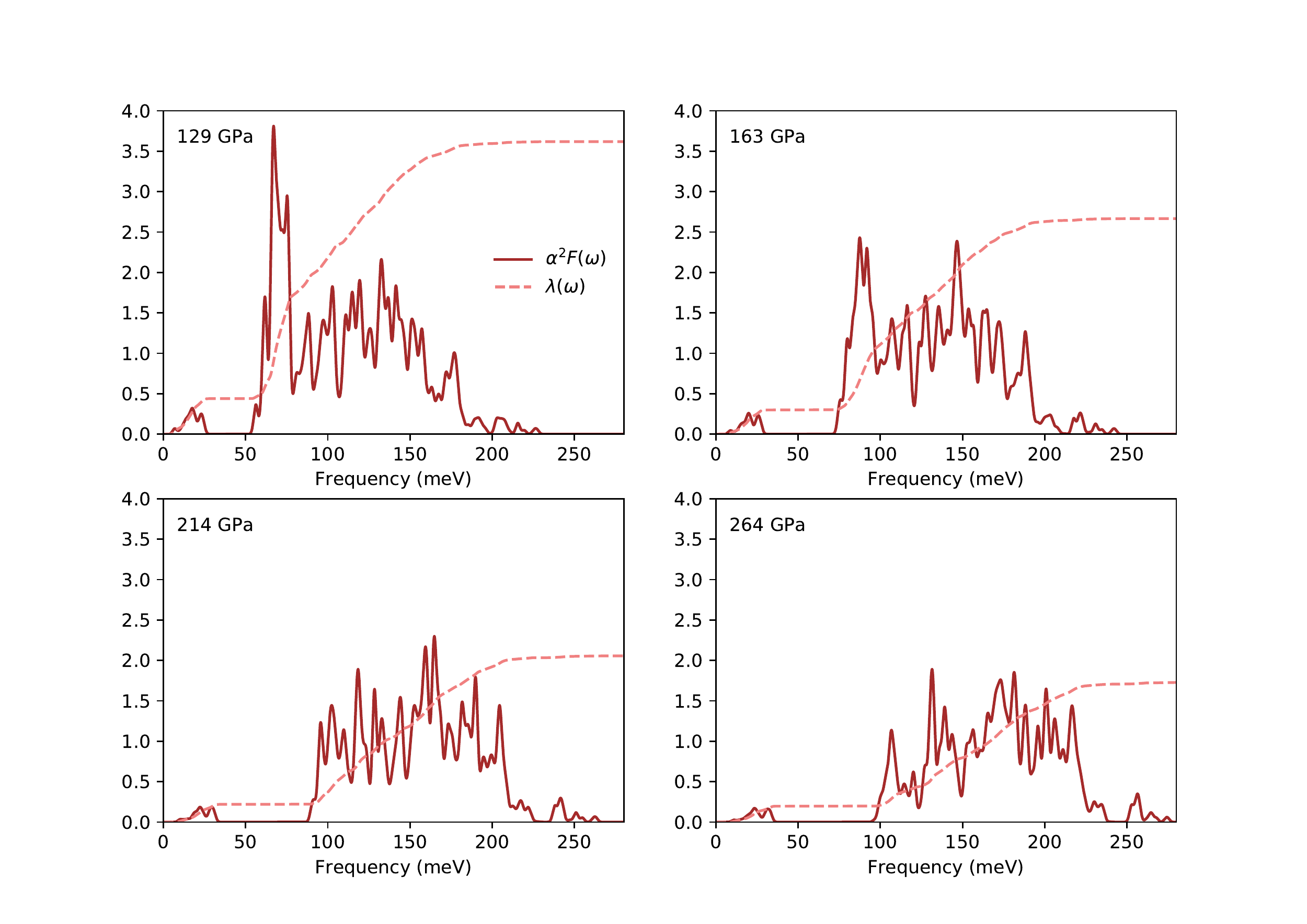}
\end{center}
\caption{~{\bf $\alpha^2F(\omega)$ for the \fcc\ phase of \laH.} 
Calculated $\alpha^2F(\omega)$ for different pressures together with the integrated electron-phonon coupling constant, which is defined as $\lambda(\omega)=2\int_0^{\omega}d\Omega \alpha^2F(\Omega)/\Omega$. The results show that optical modes, who have hydrogen character, are responsible for the large value of the electron-phonon coupling constant $\lambda$. It is worth noting, however, that acoustic modes with La character contribute between 0.2 and 0.5 to $\lambda$ and cannot be neglected to estimate \tc\ correctly.
}\label{fig:LaH10_fcc_a2f}
\end{figure*}

\newpage

\begin{figure*}[ht!]
\begin{center}
\includegraphics[width=0.7\textwidth]{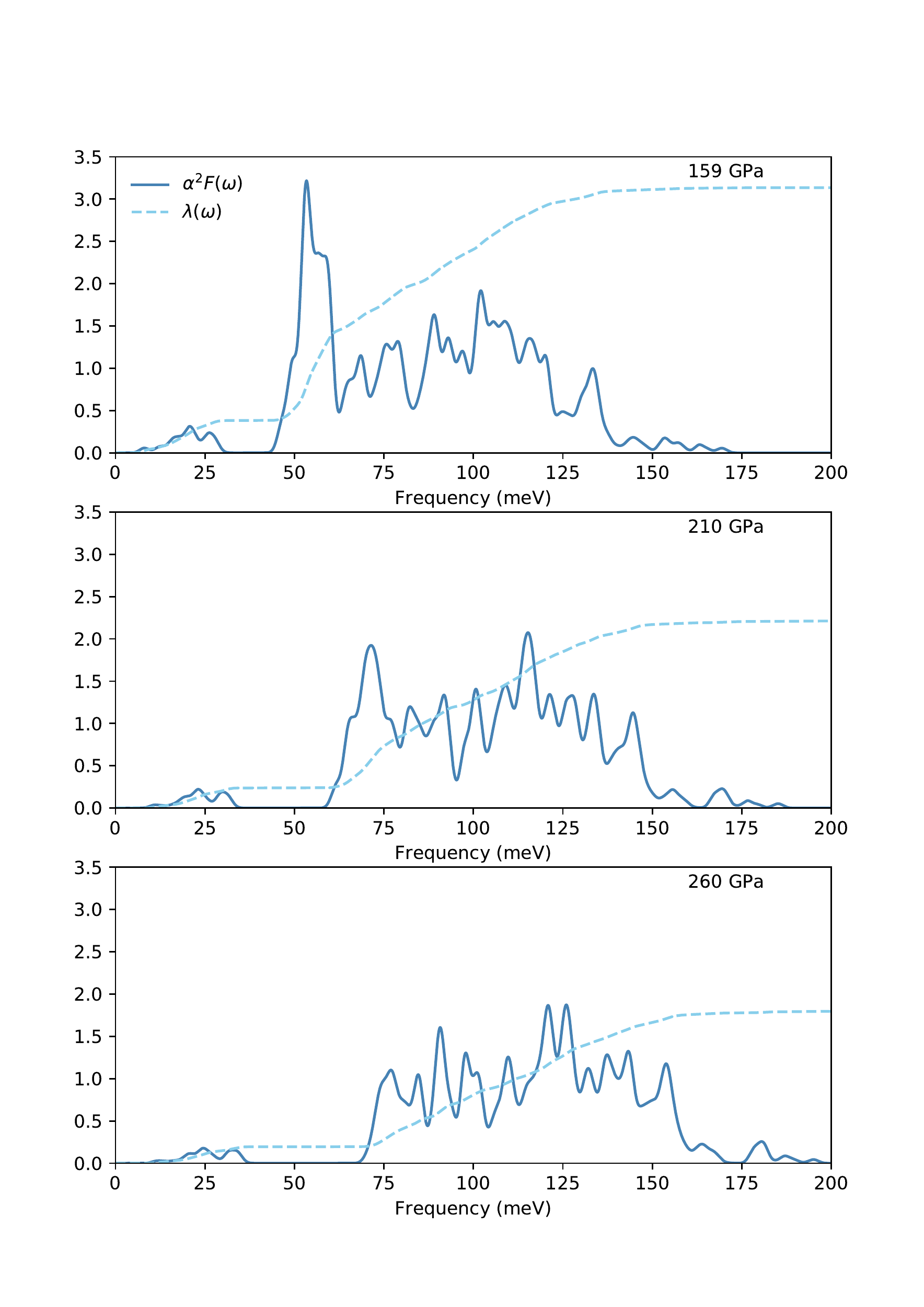}
\end{center}
\caption{~{\bf $\alpha^2F(\omega)$ for the \fcc\ phase of \laD.} 
The integrated electron-phonon coupling constant $\lambda(\omega)$ is also shown.
}\label{fig:LaD10_fcc_a2f}
\end{figure*}

\newpage

\begin{figure*}[h]
\begin{center}
\includegraphics[width=1.0\textwidth]{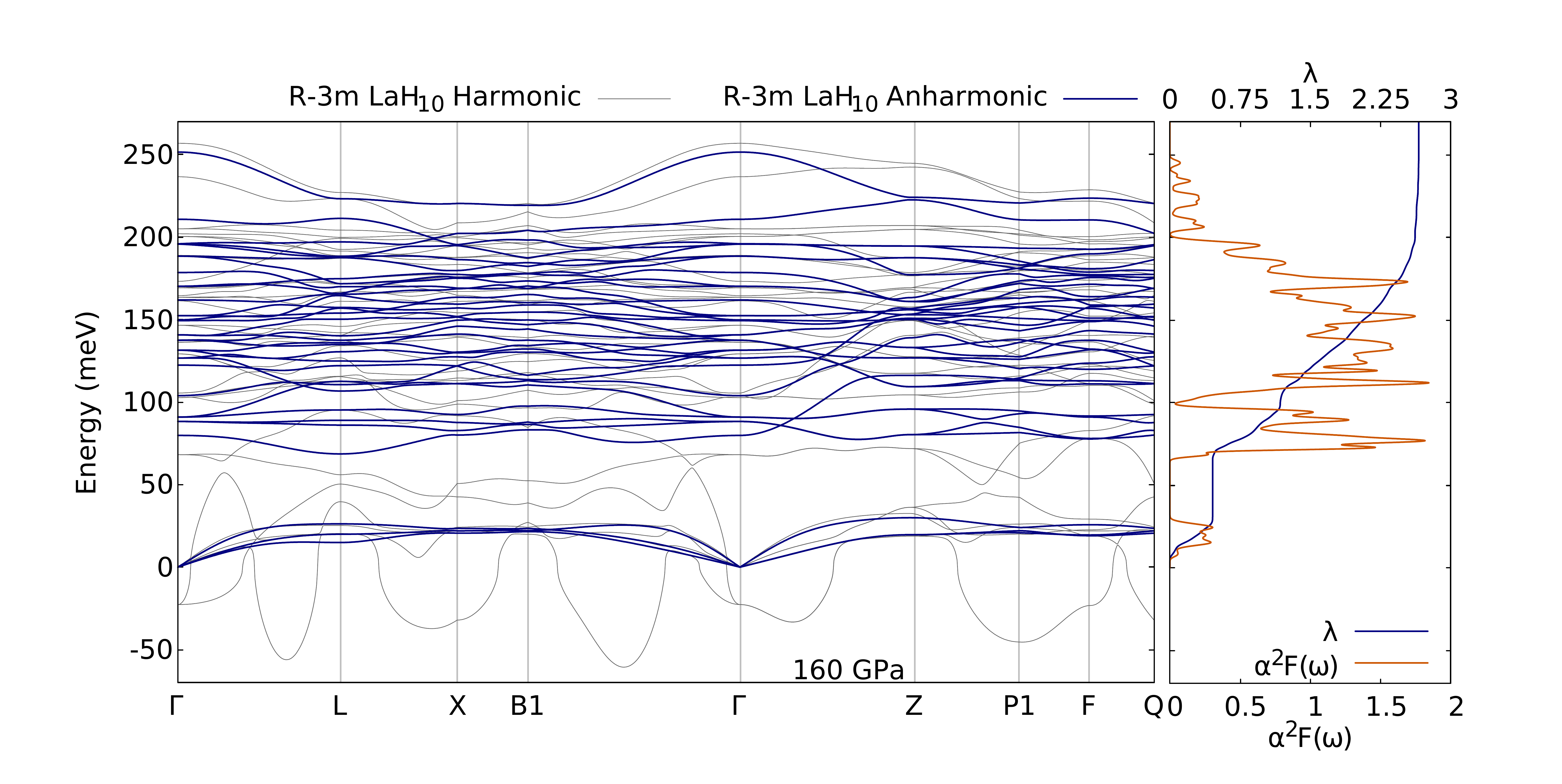}
\end{center}
\caption{~{\bf Phonon dispersion of LaH$_{10}$ on the rhombohedral phase.} 
Harmonic and anharmonic phonon spectrum keeping a 62.3\textdegree\ rhombohedral angle. The harmonic calculation is performed with the internal atomic positions that yield classical vanishing forces. The anharmonic calculation is performed after relaxing with the SSCHA the internal degrees of freedom but keeping the 62.3\textdegree\ rhombohedral angle.  
At the harmonic level there are unstable phonon modes even at $\Gamma$. Symmetry prevents the relaxation of this structure according to the unstable phonon mode at $\Gamma$. 
The harmonic phonons are calculated at a classic pressure of 150\,GPa. 
Quantum effects add an extra $\sim$10\,GPa to the pressure. 
On the right side of the figure is also shown the behavior of $\lambda(\omega)$ and $\alpha^2F(\omega)$ for the anharmonic calculation.}
\label{fig:R-3m_Ph_LaH10}
\end{figure*}

\newpage

\begin{table*}[t]
\caption{~{\bf Summary of calculated superconducting \tc.} 
Values are within different approaches ranging from empirical to fully {\it ab initio}: 
McMillan equation (\tc$^{\text Mc} _{\mu^*=0.1}$), 
Allen-Dynes modified McMillan equation (\tc$^{\text AD} _{\mu^*=0.1}$), 
anisotropic treatment of Migdal-\'Eliashberg (\tc$_{ani}^{\text ME}$) 
and SCDFT (\tcscdft). 
Values of $\lambda$ and \omlog\ are also given.}  
\begin{center}
\begin{tabular}{c| c | c | c |c| c | c | c }
 System  & Pressure (GPa) & $\lambda$  &\omlog\ (meV) &  \tc$^{\text Mc} _{\mu^*=0.1}$ (K)
 & \tc$^{\text AD} _{\mu^*=0.1}$ (K) & \tc$_{ani}^{\text ME}$ (K) &  \tcscdft\ (K)  \\ \hline
%          P(GPa)   lambda    omega     McM    McM-AD  Elias   SCDFT 
%--------|---------|----------|-------|-------|-------|-------|-------- col: 8
 \laH    &  129    &   3.62   & 76.4  & 171.8 & 252.6 & 255.3  & 230 \\ 
 \laH    &  163    &   2.67   & 96.4  & 197.1 & 247.0 & 242.8  & 225 \\
 \laH    &  214    &   2.06   & 115.5 & 196.3 & 235.9 & 237.9  & 210 \\ 
 \laH    &  264    &   1.73   & 126.6 & 189.5 & 219.2 & 216.9  & 201 \\ \hline
 %--------|--------|----------|-------|-------|-------|--------|--------- 
 \laD    &  159    &   3.14   & 63.5  & 135.0 & 184.2 & 180.4  & 171 \\
 \laD    &  210    &   2.21   & 81.7  & 145.5 & 176.5 & 172.9  & 158 \\  
 \laD    &  260    &   1.80   & 92.2  & 142.2 & 164.6 & 157.9  & 151 \\ \hline
%--------|---------|----------|-------|-------|---------|-------|--------- 
\end{tabular}
\end{center} \label{Tab:Master}
\end{table*}

\clearpage

\begin{figure*}[h]
\begin{center}
\includegraphics[width=1.0\textwidth]{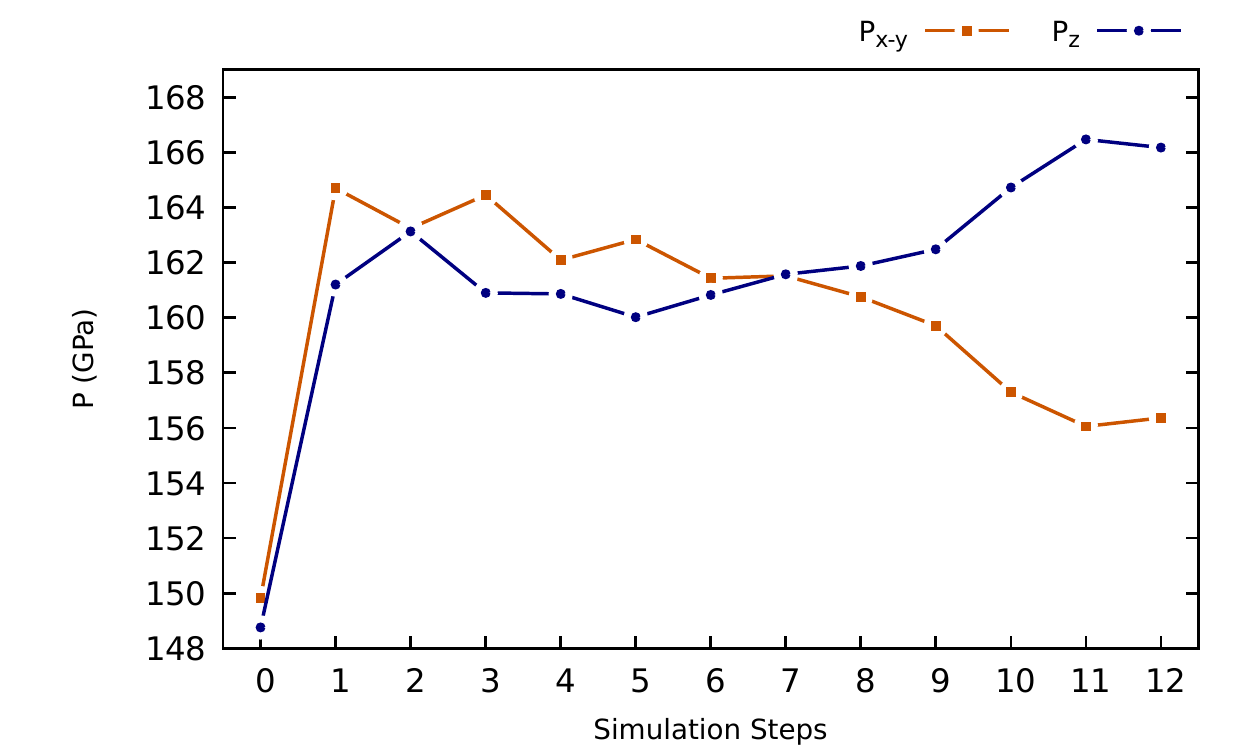}
\end{center}
\caption{~{\bf Anisotropic pressure of the $R$-$3m$ phase of LaH$_{10}$ in a fixed-cell quantum relaxation.}
Pressure along the different Cartesian directions during the SSCHA relaxation of the internal parameters keeping the rhombohedral angle at 62.3\textdegree\ fixed.  At  step 0 the pressure reported is obtained directly from $V(\mathbf{R})$, neglecting quantum effects. It is isotropic within one GPa of difference between the $x-y$ and $z$ directions. At the other steps it is calculated from the quantum $E(\boldsymbol{\mathcal{R}})$ and along the minimization it becomes anisotropic. When the minimization stops at step 12, i.e., the internal coordinates are at the minimum of the $E(\boldsymbol{\mathcal{R}})$ for this lattice, the stress anisotropy between $z$ and $x-y$ directions is about 6\%. This clearly indicates that quantum effects want to relax the crystal lattice, in particular, since $P_z$ is larger, by reducing the rhombohedral angle.  It is worth noting that quantum effects approximately increase the total pressure by 10 GPa, which is calculated as $P=(P_x+P_y+P_z)/3$. The initial cell parameters before the minimization are $a = 3.5473398$ \AA\ and $\alpha = 62.34158$ \degree\ . The initial values of the free Wyckoff parameters, which yield classical vanishing forces and a 150 GPa isotropic stress, are $\epsilon_a = 0.26043$, $\epsilon_b = 0.09950$, $\epsilon_x = 0.10746$ and $\epsilon_y = 0.12810$. Check the Extended Data Table \ref{tab:LaH10_Wyckoff_tab} for more details.
}
\label{fig:R-3m_IntRel_Pressure_LaH10}
\end{figure*}

\newpage

\begin{figure*}[h]
\begin{center}
\includegraphics[width=0.49\textwidth]{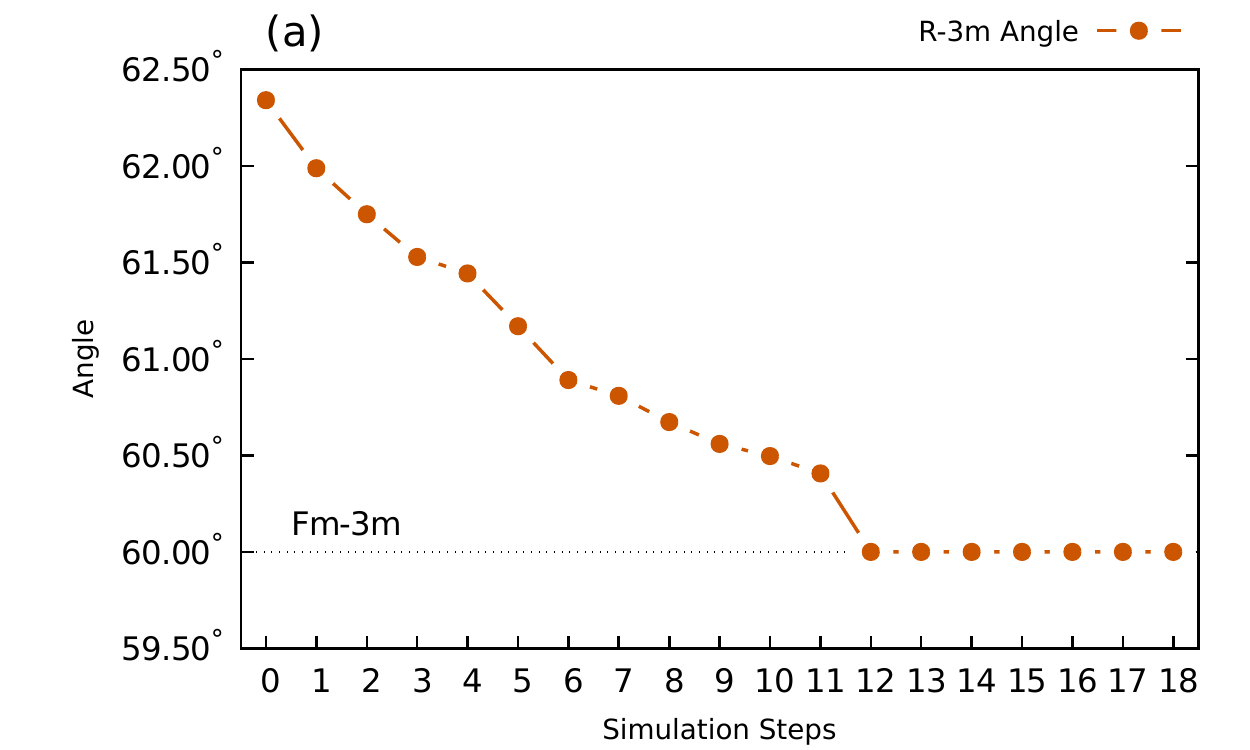} %
\includegraphics[width=0.49\textwidth]{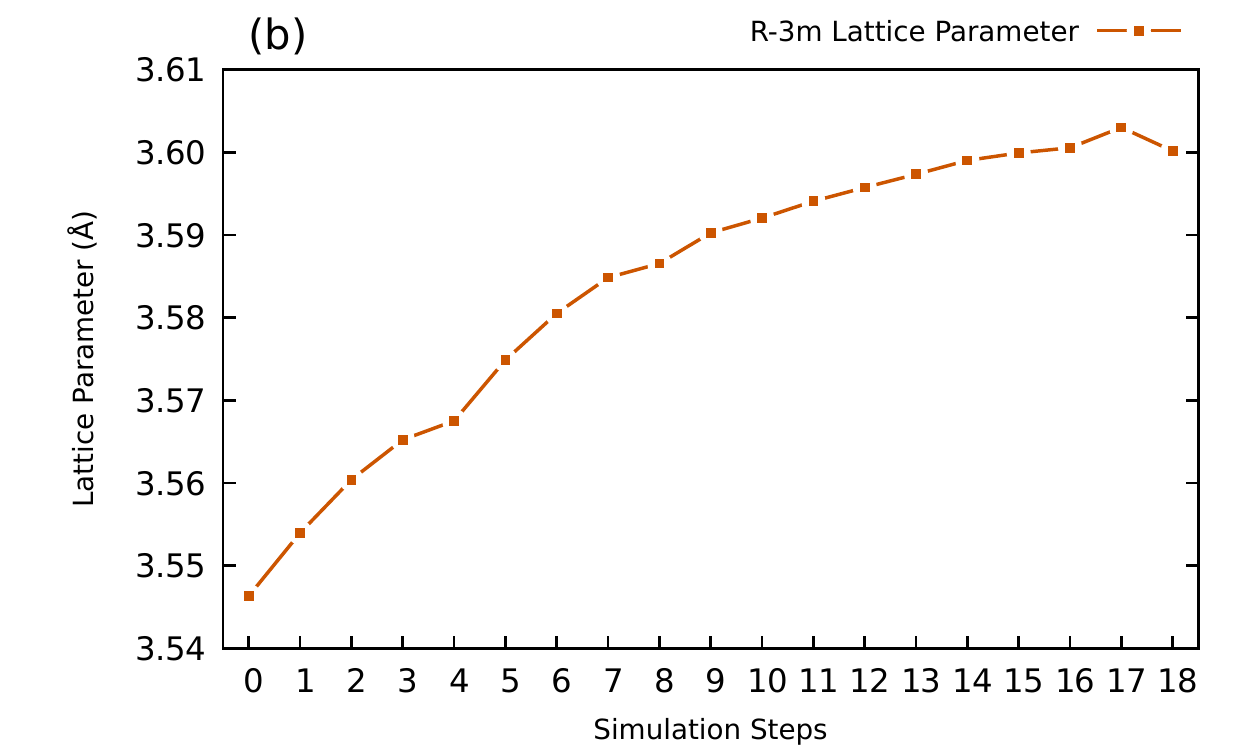} %
\includegraphics[width=0.49\textwidth]{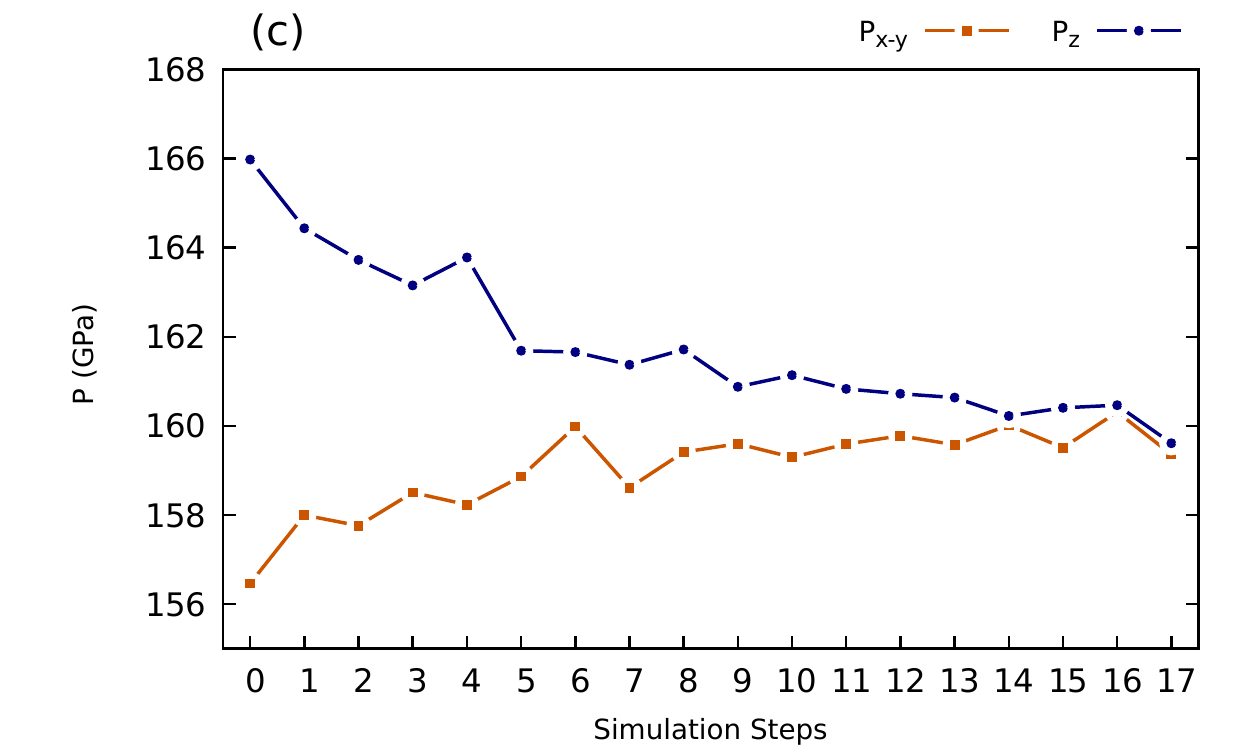} %
\includegraphics[width=0.475\textwidth]{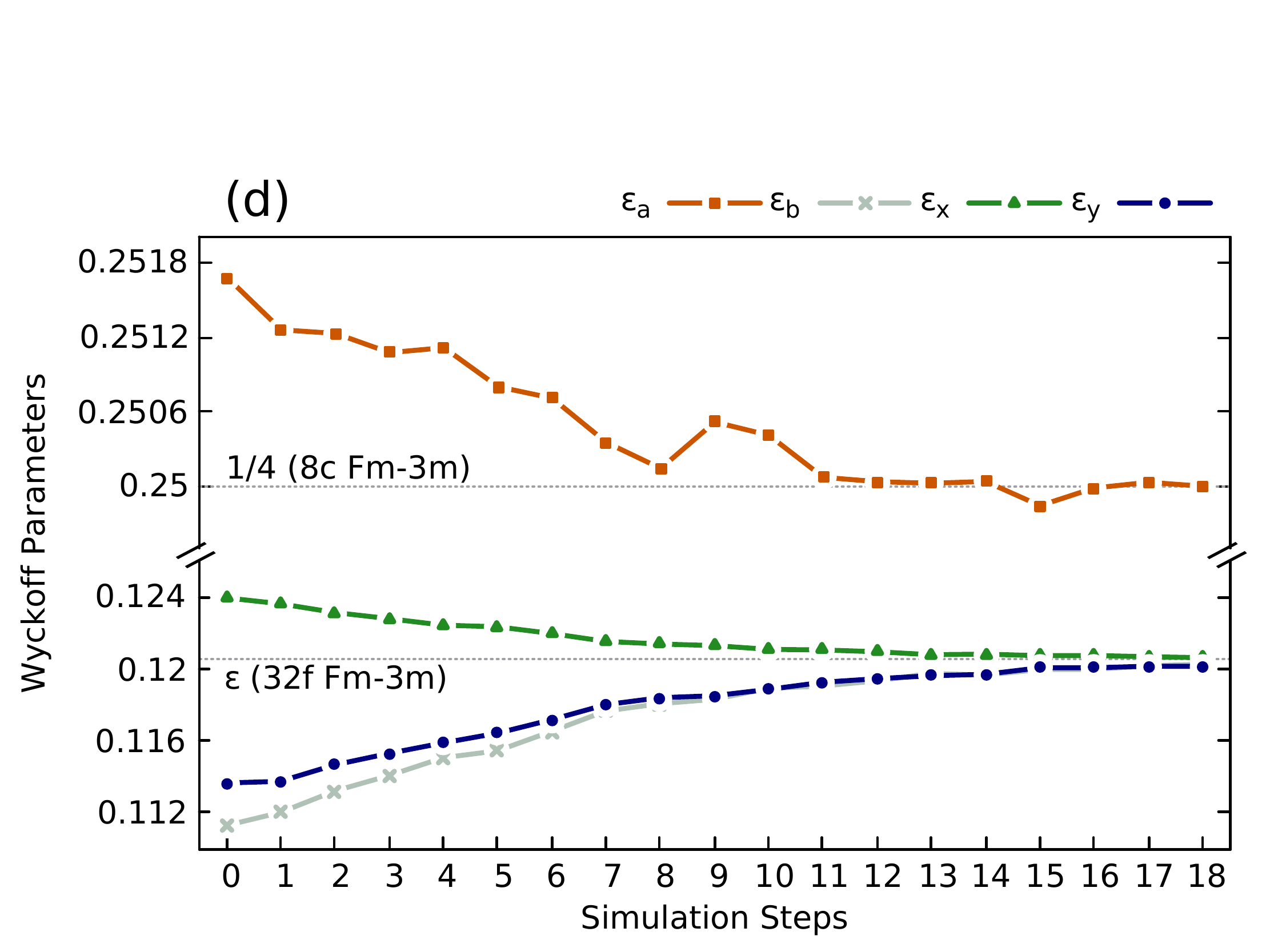} 

\end{center}
\caption{~{\bf Details on the $R$-$3m$ LaH$_{10}$ cell relaxation including quantum effects.} The initial point for the relaxation is the output from the previous internal relaxation with fixed angle presented in Extended Figure \ref{fig:R-3m_IntRel_Pressure_LaH10}. The $R$-$3m$ phase in the rhombohedral description is described by three vectors of the same length ($a = b = c$) and by the angle between them ($\alpha=\beta=\gamma$). In panel \textbf{(a)} we show the evolution of the rhombohedral angle and in panel \textbf{(b)} the evolution of the rhombohedral lattice parameter ($|\mathbf{a}| = |\mathbf{b}| = |\mathbf{c}|$). The evolution of the stress tensor in the quantum SSCHA minimization is shown in panel \textbf{(c)}. It is clear that in the end of the minimization the structure obtains a 60\textdegree\ angle, which matches the angle of a fcc\ lattice, and that in this case the stress is isotropic. In panel \textbf{(d)} we show the evolution of the Wyckoff positions in the minimization and we compare it with those of the \fcc. The occupied Wyckoff positions for both $R$-$3m$ LaH$_{10}$ and \fcc\ LaH$_{10}$ are summarized in the Extended Data Table \ref{tab:LaH10_Wyckoff_tab}.  Here, it can be seen the evolution of $\epsilon_a$, $\epsilon_b$, $\epsilon_x$, and $\epsilon_y$ parameters in the minimization. The atoms in the first set of $6c$ positions approach the $8c$ Wyckoff site of the \fcc, while the atoms in the second  set of $6c$ positions and those in $18h$ sites approach the atoms in the $32f$ Wyckoff site of the \fcc, where $\epsilon = 0.12053$.}
\label{fig:R-3m_CellRelax_LaH10}
\end{figure*}

\begin{table*}[t]
    \caption{\textbf{$R$-$3m$ LaH$_{10}$ and $Fm$-$3m$ LaH$_{10}$ Wycokff positions.} The table summarizes the occupied Wyckoff positions for the two structures. We describe the Wyckoff positions in crystal coordinates so that the  $\left[x,y,z\right]$ coordinate should be understood as a $x\mathbf{a} + y \mathbf{b} + z \mathbf{c}$ atomic position with $\mathbf{a}$, $\mathbf{b}$, $\mathbf{c}$ the lattice vectors. For the $R$-$3m$ phase we use the rhombohedral lattice (R), where the three lattice vectors have the same length ($a = b = c$) and the angle between them is the same ($\alpha=\beta=\gamma$). The \fcc\ phase is described both in this rhombohedral description (R) and, for comparison, in the standard cubic conventional lattice (C). In the \fcc\ the lanthanum atom is described by the \textbf{4b} sites, two hydrogen atoms occupy the \textbf{8c} sites, and the remaining 8 hydrogen atoms occupy the \textbf{32f} sites. Most of the atomic positions are fixed by symmetry and overall the \fcc\ structure can be described by one single free parameter ($\epsilon$). In the $R$-$3m$ the lanthanum atom is locked in the \textbf{3b} sites, two pairs of hydrogen atoms  occupy the \textbf{6c} sites, and the remaining 6 hydrogen atoms occupy the \textbf{18h} sites. In this case symmetry allows for more freedom, and overall the structure of the $R$-$3m$ phase can be described by 4 free parameters ($\epsilon_a$, $\epsilon_b$, $\epsilon_x$ and $\epsilon_y$).}
\begin{center}
    \begin{tabular}{l l|l l|l l}
 \multicolumn{2}{c|}{ $Fm$-$3m$ (C) } & %
 \multicolumn{2}{c|}{ $Fm$-$3m$ (R) } & %
 \multicolumn{2}{c}{ $R$-$3m$ (R) } \\ [1ex]
 \hline
 %%%%%%%%%%%%%
 &&&&&\\
 1 La \textbf{4b} & $\left[\frac{1}{2},\frac{1}{2},\frac{1}{2}\right]$ & 1 La \textbf{4b} &$\left[\frac{1}{2},\frac{1}{2},\frac{1}{2}\right]$ & 1 La \textbf{3b} &$\left[\frac{1}{2},\frac{1}{2},\frac{1}{2}\right]$\\
 %%%%%%%%%%%%%
 &&&&&\\
 %%%%%%%%%%%%%%%%%%%%%%%%%%%%%%%%
2 H \textbf{8c} & $\left[\frac{1}{4},\frac{1}{4},\frac{1}{4}\right]$ & 2 H \textbf{8c} & $\left[\frac{1}{4},\frac{1}{4},\frac{1}{4}\right]$ & 2 H \textbf{6c} &  [$\epsilon_a,\epsilon_a,\epsilon_a$] \\[0.5ex]
 %%%%%%%%%%%%%%%%%%%%%%%%%%%%%%%%
 & $\left[\frac{3}{4},\frac{3}{4},\frac{3}{4}\right]$ &  & $\left[\frac{3}{4},\frac{3}{4},\frac{3}{4}\right]$ &     & [-$\epsilon_a$,-$\epsilon_a$,-$\epsilon_a$]\\
%%%%%%%%%%%%%
 &&&&&\\
 %%%%%%%%%%%%%%%%%%%%%%%%%%%%%%%%%%%%
8 H \textbf{32f} & [$\epsilon$,$\epsilon$,$\epsilon$] & 8 H \textbf{32f} & [$\epsilon$,$\epsilon$,$\epsilon$] &  2 H \textbf{6c} & [$\epsilon_b$,$\epsilon_b$,$\epsilon_b$] \\[0.5ex]
 %%%%%%%%%%%%%%%%%%%%%%%%%%%%%%%%%%%%
 & [-$\epsilon$,-$\epsilon$,-$\epsilon$] &  & [-$\epsilon$,-$\epsilon$,-$\epsilon$] &   & [-$\epsilon_b$,-$\epsilon_b$,-$\epsilon_b$] \\[0.5ex]
  %%%%%%%%%%%%%%%%%%%%%%%%%%%%%%%%%%%%
 & [$\epsilon$,$\epsilon$,-$\epsilon$] &  & [-$\epsilon$,-$\epsilon$,3$\epsilon$] & 6 H \textbf{18h}& [-$\epsilon_x$,-$\epsilon_x$,$3\epsilon_y$] \\[0.5ex]
   %%%%%%%%%%%%%%%%%%%%%%%%%%%%%%%%%%%%
 & [$\epsilon$,-$\epsilon$,$\epsilon$] &  & [-$\epsilon$,3$\epsilon$,-$\epsilon$] &  & [-$\epsilon_y$,$3\epsilon_x$,-$\epsilon_x$] \\[0.5ex]
    %%%%%%%%%%%%%%%%%%%%%%%%%%%%%%%%%%%%
 & [-$\epsilon$,$\epsilon$,$\epsilon$] &  & [3$\epsilon$,-$\epsilon$,-$\epsilon$] &  & [$3\epsilon_y$,-$\epsilon_x$,-$\epsilon_x$] \\[0.5ex]
   %%%%%%%%%%%%%%%%%%%%%%%%%%%%%%%%%%%%
 & [-$\epsilon$,-$\epsilon$,$\epsilon$] &  & [$\epsilon$,$\epsilon$,-3$\epsilon$] &  & [$\epsilon_x$,$\epsilon_x$,-$3\epsilon_y$] \\[0.5ex]
   %%%%%%%%%%%%%%%%%%%%%%%%%%%%%%%%%%%%
 & [-$\epsilon$,$\epsilon$,-$\epsilon$] &  & [$\epsilon$,-3$\epsilon$,$\epsilon$] &  & [$\epsilon_x$,-$3\epsilon_y$,$\epsilon_x$] \\[0.5ex]
    %%%%%%%%%%%%%%%%%%%%%%%%%%%%%%%%%%%%
 & [$\epsilon$,-$\epsilon$,-$\epsilon$] &  & [-3$\epsilon$,$\epsilon$,$\epsilon$] &  & [-$3\epsilon_y$,$\epsilon_x$,$\epsilon_x$] \\[1ex]
 \hline
    
\end{tabular}
\end{center}
    \label{tab:LaH10_Wyckoff_tab}
\end{table*}

\begin{table*}[t]
    \caption{\textbf{Crystal structure details for relevant phases}. 
    Lattice parameters and atomic coordinates for 
    LaH$_{10}$ ($Immm$) and LaH$_{10}$ ($C2$) at 150\,GPa 
    and LaH$_{11}$ $P4/nmm$ at 100\,GPa. These pressures are estimated classically. The positions below give vanishing forces at classical level.}
\begin{center}
    \begin{tabular}{p{3cm} p{4.5cm} l l l}
 \multicolumn{1}{l}{ Composition (Space group) } & %
 \multicolumn{1}{l}{ Lattice parameters } & %
 \multicolumn{3}{l}{ Wyckoff positions  } \\ [1ex]
 \hline
 %%%%%%%%%%%%%
  & & & & \\[0.5ex]
 LaH$_{10}$ ($Immm$) & $a =  3.58303$ \AA & La  & \textbf{2c}  & [0.50000, 0.50000, 0.00000]\\[0.5ex]
                   & $b =  3.61834$ \AA & H   & \textbf{8m}  & [0.75841, 0.00000, 0.11649]\\[0.5ex]
                   & $c =  5.08749$ \AA& H   & \textbf{8l}  & [0.00000, 0.75742, 0.87548]\\[0.5ex]
                   &                & H   & \textbf{4j}  & [0.50000, 0.00000, 0.74572]\\[2ex]

LaH$_{10}$ ($C2$)  & $a =  6.15468$ \AA     & La & \textbf{4c}  & [0.49244, 0.00070, 0.25292]\\[0.5ex]
                   & $b =  3.60628$ \AA     & H  & \textbf{4c}  & [0.13978, 0.24567, -0.05243]\\[0.5ex]
                   & $c =  7.23776$ \AA     & H  & \textbf{4c}  & [0.09798, 0.24122, 0.45027]\\[0.5ex]
                   & $\beta =  55.71434$\textdegree & H  & \textbf{4c}  & [0.36015, 0.25590, 0.05238]\\[0.5ex]
                   &                     & H  & \textbf{4c}  & [0.40204, 0.26021, 0.54971]\\[0.5ex]
                   &                     & H  & \textbf{4c}  & [-0.09751, 0.00051, -0.05100]\\[0.5ex]
                   &                     & H  & \textbf{4c}  & [0.86810, 0.00071, 0.43706]\\[0.5ex]
                   &                     & H  & \textbf{4c}  & [0.88713, 0.00076, 0.69398]\\[0.5ex]
                   &                     & H  & \textbf{4c}  & [0.87083, 0.00068, 0.19089]\\[0.5ex]
                   &                     & H  & \textbf{4c}  & [0.73058, 0.00043, 0.88088]\\[0.5ex]
                   &                     & H  & \textbf{4c}  & [0.76156, 0.00071, 0.36763]\\[2ex]

LaH$_{11}$ ($P4/nmm$)& $a =  3.87435$  \AA    & La & \textbf{2c}  & [0.25000, 0.25000, 0.78577]\\[0.5ex]
                   & $b =  3.87435$ \AA     & H  & \textbf{4e}  & [0.00000, 0.00000, 0.50000]\\[0.5ex]
                   & $c =  5.27636$ \AA     & H  & \textbf{8i}  & [0.25000, -0.02052, 0.17824]\\[0.5ex]
                   &                     & H  & \textbf{8i}  & [0.25000, 0.55418, 0.35160]\\[0.5ex]
                   &                     & H  & \textbf{2a}  & [0.75000, 0.25000, 0.00000]\\[2ex]
 \hline
    
\end{tabular}
\end{center}
    \label{tab:Wyckoff_tab2}
\end{table*}

\newpage

\begin{figure*}[h]
\begin{center}
\includegraphics[width=0.8\textwidth]{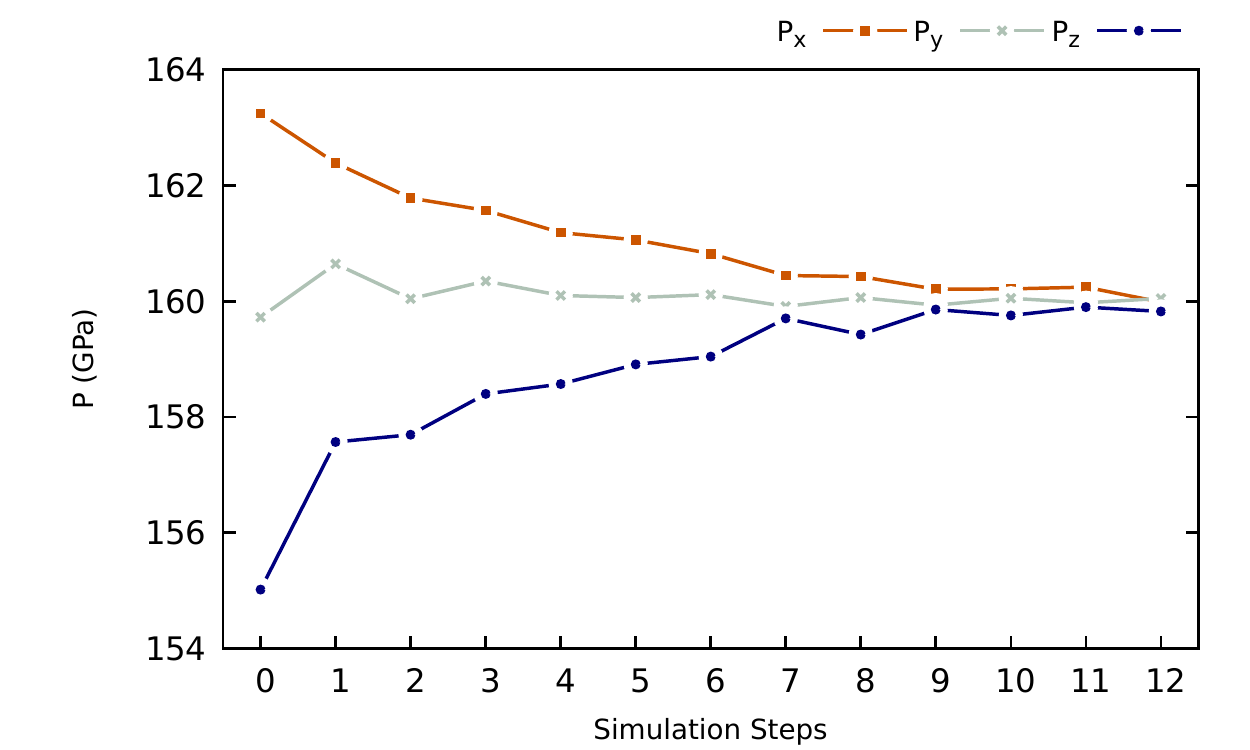}
\end{center}
\caption{~{\bf Anisotropic pressure of the $C2$ phase of LaH$_{10}$ in a cell quantum relaxation.} The figure shows the pressures along the different Cartesian directions during the SSCHA cell minimization. The target pressure for this minimization is 160 GPa. At the end of the minimization the isotropy of the stress tensor is recovered. A symmetry analysis performed on the structure at the end of the minimization confirms the $C2$ LaH$_{10}$ evolves in the \fcc\ LaH$_{10}$. The initial values $P_x = 163.2$ (GPa), $P_y = 159.7$(GPa), $P_z = 155.0$(GPa) are obtained by an atomic internal relaxation performed using the SSCHA with fixed cell.}
\label{fig:C2_Relax_Pressure_LaH10}
\end{figure*}

\newpage

\begin{figure*}[ht!]
\begin{center}
\includegraphics[width=1.0\textwidth]{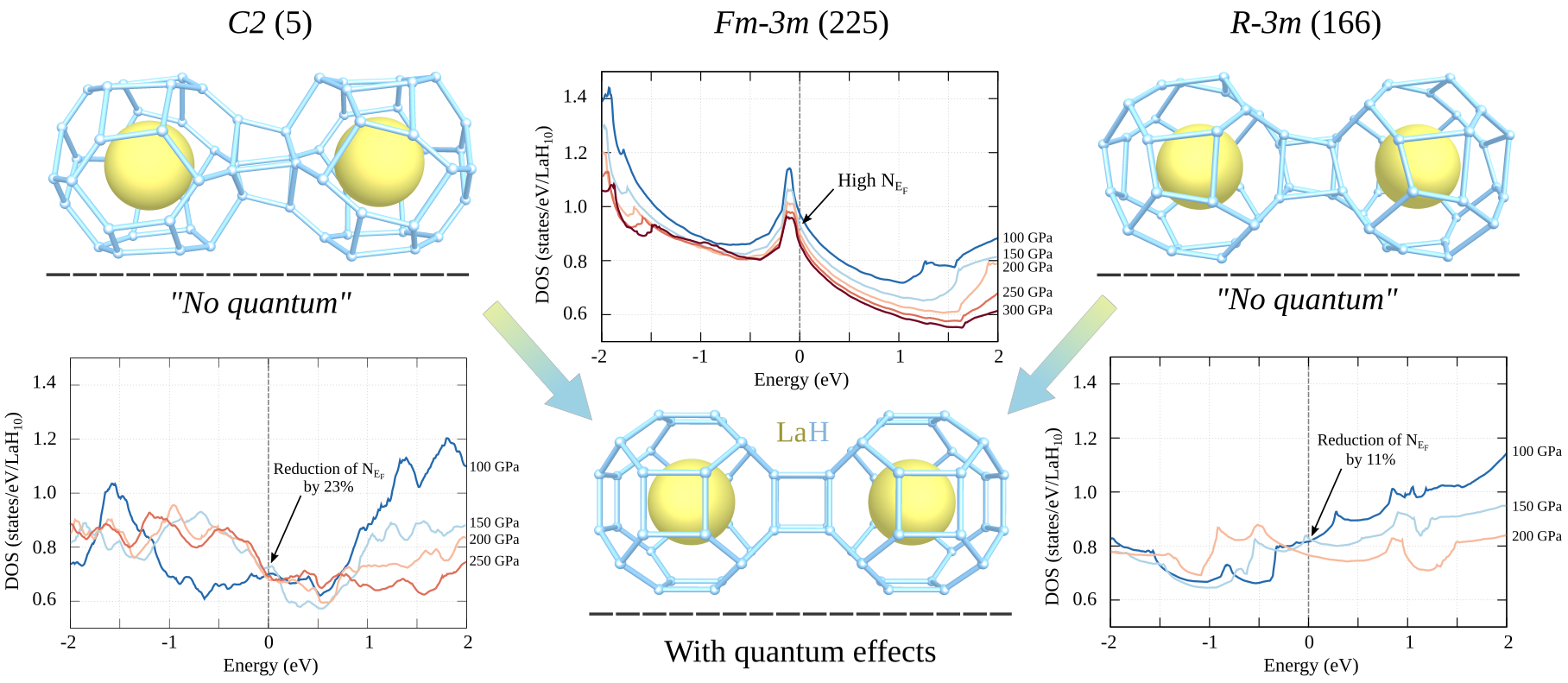}
\end{center}
\caption{~{\bf SSCHA minimization on LaH$_{10}$ and DOS.} 
Top figures: two initial structures ($C2$ and $R$-$3m$) low enthalpy, considered in our SSCHA simulations. 
When considering quantum effects both structures evolve towards the \fcc\ structure. 
Corresponding total electronic density of states (DOS) as a function of pressure are 
plotted for each structure (for comparison in the same energy scale).  
Highly symmetric motif ($Fm$-$3m$) maximizes $N_{E_{F}}$, 
while in distorted structures ($R$-$3m$ and $C2$) the occupation at the Fermi level 
is reduced by more than 23~\% for $C2$ and by 11~\% for $R$-$3m$, both comparison at 150\,GPa w.r.t. \fcc .
Classical pressures are appended for comparison in DOS panels. 
Note that DOS shape is also strongly modified.}\label{fig:Ini-Fin_structures} 
\end{figure*}

\newpage

\begin{figure*}[ht!]
\begin{center}
\includegraphics[width=1.0\textwidth]{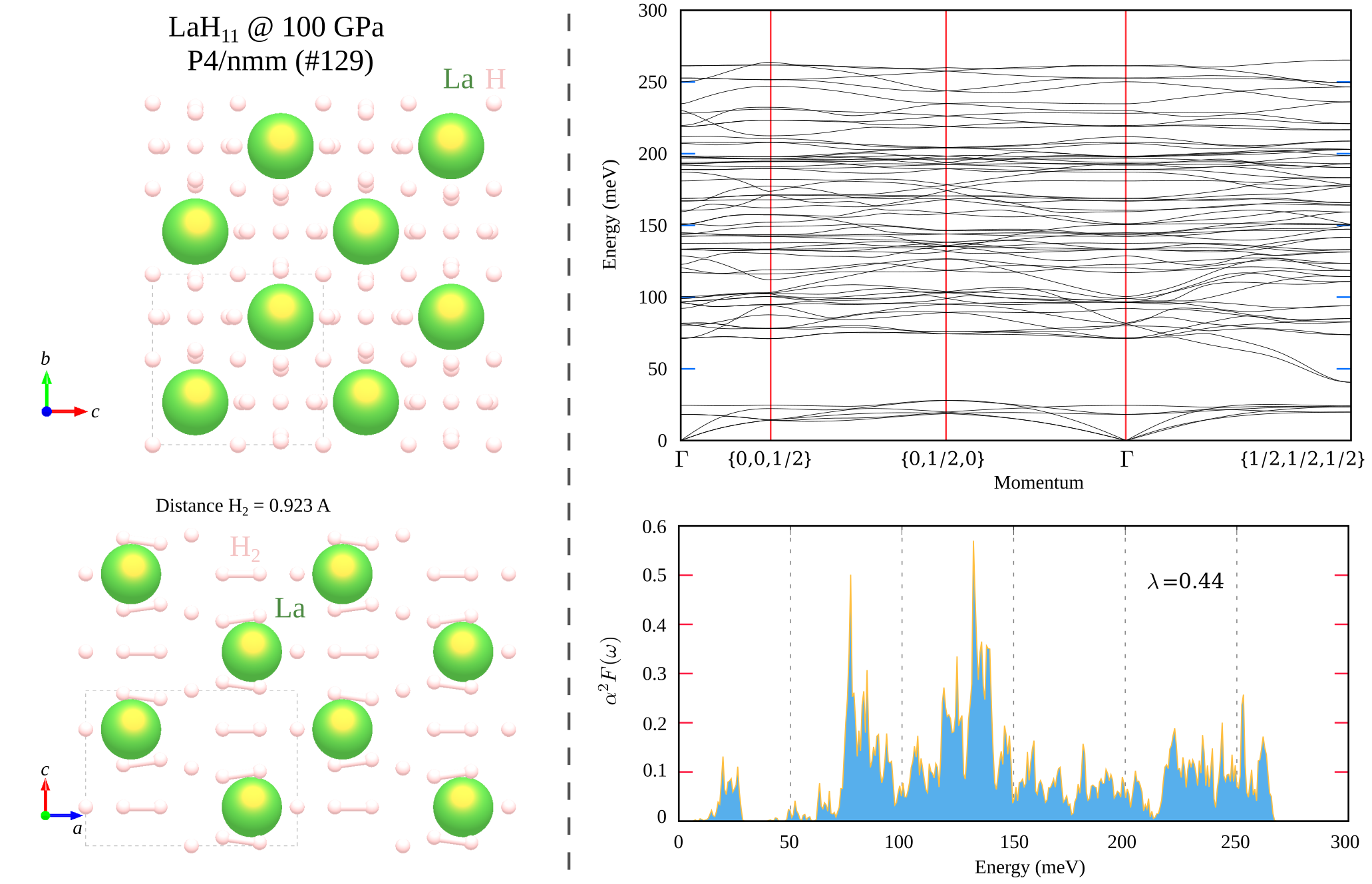}
\end{center}
\caption{~{\bf Details on LaH$_{11}$.} 
 Left: $P4/nmm$ crystal structure of LaH$_{11}$ at 100\,GPa thermodynamically stable in the convex hull.
 Right top: harmonic phonons dispersion along momentum space for this composition: it is dynamically stable. 
 Right bottom: superconducting \'Eliashberg spectrum function ($\alpha^2 F(\omega)$) 
 calculated for this composition at the pressure indicated with harmonic phonons.
   The estimated \tc\ with Allen-Dynes formula ($\mu^*$=0.1) 
 is $\sim$7\,K at 100\,GPa (harmonic phonons).}\label{fig:LaH11}
\end{figure*}

%===============
\end{document}